\documentclass[12pt]{iopart}

\expandafter\let\csname equation*\endcsname\relax
\expandafter\let\csname endequation*\endcsname\relax

\usepackage[english]{babel}
\usepackage{microtype}

\usepackage{graphicx}
\usepackage{cite}
\usepackage{physics}
\usepackage{iopams}

\usepackage[table]{xcolor}
\usepackage{tabularx}
\usepackage{makecell}
\usepackage{nicematrix}
\usepackage{pifont}

\newcommand{\cmark}{\ding{51}}
\newcommand{\xmark}{\ding{55}}

\newcommand{\Prb}{\mathbb{P}}
\newcommand{\Exp}{\mathbb{E}}
\newcommand{\Orb}{\mathcal{O}}
\newcommand{\im}{\mathrm{i}}
\DeclareMathOperator{\len}{len}

\newcommand{\keywords}[1]{\noindent{\it Keywords\/}: #1}

\begin{document}

\title[Weak chaos for open intermittent dynamical systems?]{Does an intermittent dynamical system remain (weakly) chaotic after drilling in a hole?}

\author{Samuel Brevitt$^1$ and Rainer Klages$^{1,2}$}

\address{$^1$ Centre for Complex Systems, School of Mathematical Sciences, Queen Mary University of London, Mile End Road, London E1~4NS, UK}
\address{$^2$ London Mathematical Laboratory, 8 Margravine Gardens, London W6~8RH, UK}

\ead{s.brevitt@qmul.ac.uk}

\begin{abstract}
Chaotic dynamical systems are often characterised by a positive
Lyapunov exponent, which signifies an exponential rate of separation
of nearby trajectories. However, in a wide range of so-called weakly
chaotic systems, the separation of nearby trajectories is
sub-exponential in time, and the Lyapunov exponent vanishes. When a
hole is introduced in chaotic systems, the positive Lyapunov exponents
on the system's fractal repeller can be related to the generation of
metric entropy and the escape rate from the system. The escape rate,
in turn, cross-links these two chaos properties to important
statistical-physical quantities like the diffusion
coefficient. However, no suitable generalisation of this escape rate
formalism exists for weakly chaotic systems. In our paper we show that
in a paradigmatic one-dimensional weakly chaotic iterated map, the
Pomeau-Manneville map, a generalisation of its Lyapunov exponent (which we
call `stretching') is completely suppressed in
the presence of a hole. This result is based on numerical evidence and
a corresponding stochastic model. The correspondence between map and
model is tested via a related partially absorbing map. We examine the
structure of the map's fractal repeller, which we reconstruct via a
simple algorithm. Our findings are in line with rigorous mathematical
results concerning the collapse of the system's density as it evolves
in time. We also examine the generation of entropy in the open map,
which is shown to be consistent with the collapsed stretching. As a
result, we conclude that no suitable generalisation of the escape rate
formalism to weakly chaotic systems can exist.
\end{abstract}

\par
\submitto{\NJP}


\section{Introduction}

Dynamical systems in nature are typically not isolated, or closed, but
instead are open, fostering complex interactions with their
environments by exchanging matter, heat, or information
\cite{ScSa05}. Stochastic theory models these situations by
supplementing a given stochastic dynamics with appropriate boundary
conditions, which for open systems are absorbing boundaries. This
setting yields paradigmatic scenarios like first passage problems
\cite{Red01,BMS13}, for which there exist famous mathematical results
like the Sparre Andersen theorem \cite{Maj10}. First passage problems
have in turn wide applications across all fields of science, as in
cold atoms, chemical reactions, climate, foraging, finance, and
computer science \cite{VLRS11,BeVo14,MOR14,PBL19,GMO24}.

However, open systems have not only been studied by stochastic methods
but also in deterministic dynamical systems theory
\cite{gaspard_chaos_1998,LaTe11,APT13,BBF14}. Prominent open chaotic dynamical
systems that have been investigated experimentally are ultracold
atoms, confined by beams of light to pre-defined geometries with a
hole on the boundary through which particles can escape
\cite{MCR01,FKCD01}. These open Hamiltonian dynamical systems have
also been studied theoretically, leading to the prediction of
universal laws for their escape dynamics
\cite{BaBe90,CrKe08,Vene09}. Simplified versions of open deterministic
dynamics are in turn amenable to proofs of particular mathematical
ergodic and dynamical systems properties
\cite{BuDe05b,DeYo06,bunimovich_where_2011}.

Another layer of results was added to the field of open dynamical
systems at the beginning of the 1990s, when stochastic and
deterministic approaches were combined within the framework of
nonequilibrium statistical physics by means of the escape rate
formalism to transport \cite{GN,GD1,GD2}, explaining the origin of
irreversible transport in time-reversible deterministic dynamical
systems by the chaotic and fractal properties of their associated
nonequilibrium steady states \cite{gaspard_chaos_1998,dorfman_introduction_1999,Voll02,Kla06}. However,
this approach only worked for well-behaved types of stochastic and
deterministic dynamics, namely ones exhibting `normal' transport
properties like Brownian motion-type diffusion, where the mean square
displacement of an ensemble of particles grows linearly in time. In
the stochastic world normal diffusion is typically generated by
Markovian, Gaussian dynamics like simple random walks, Wiener or
Ornstein-Uhlenbeck processes \cite{vK,Risk,Gard09}. These processes
can be related by probabilistic coarse graining \cite{castiglione_chaos_2008} to
deterministic dynamical systems that are `chaotic' in the sense of
exhibiting a positive Lyapunov exponent, which for sake of clarity in
the following we denote as `strongly chaotic'. Strongly chaotic
dynamical systems typically generate normal diffusion in suitable
settings \cite{gaspard_chaos_1998,dorfman_introduction_1999,Kla06}. But it is well-known that there are
other dynamical systems displaying irregular dynamics characterised by
zero Lyapunov exponents, called weakly chaotic
\cite{ZaUs01,Gala03,ArCr05,Kla06,klages_weak_2013}. Widely studied
examples are polygonal billiards \cite{Kla06,Gut96} and
Pomeau-Manneville maps, where the latter have been introduced to model
intermittency in turbulence
\cite{manneville_intermittency_1979,manneville_intermittency_1980,pomeau_intermittent_1980}.
These types of dynamics correspond in a probabilistic description to
more non-trivial, non-Markovian and/or non-Gaussian stochastic
processes like L\'evy flights and walks \cite{ZDK15}, fractional
Brownian motion \cite{embrechts2009selfsimilar}, or continuous time
random walks \cite{KlSo11}, to just name a few generic examples. All
these deterministic and stochastic processes generate what became
known as anomalous diffusion, where the mean square displacement grows
nonlinearly in time, either sub- or superdiffusively (corresponding to
a sub- or superlinear spreading of particles with time, respectively)
\cite{BoGe90,MeKl00,klages_anomalous_2008,MJCB14}.

Characterising these different types of dynamics by their escape
properties if systems are open on bounded domains, it was found that
for strongly chaotic dynamics, (respectively, Markovian, Gaussian
stochastic processes), the number of particles typically decays
exponentially in time \cite{Schu,Ott,gaspard_chaos_1998,dorfman_introduction_1999,Kla06}, while for
weakly chaotic dynamics, (respectively, non-Markovian, non-Gaussian
stochastic processes), particle escape often exhibits power-law decay
\cite{BaBe90,ZaUs01,CrKe08,Vene09}. Importantly, for strongly chaotic
dynamical systems, an escape rate formula could be derived expressing
the exponential escape rate in terms of the difference between the
positive Lyapunov exponents and the metric entropy on the emerging
fractal repeller
\cite{kadanoff_escape_1984,kantz_repellers_1985,bohr_entropy_1987}. This
result is in turn a generalisation of the famous Pesin identity, which
holds for strongly chaotic closed dynamical systems
\cite{ER,gaspard_chaos_1998,dorfman_introduction_1999,Kla17}. That strongly chaotic open dynamics
exhibits the same type of escape as corresponding Markovian Gaussian
open stochastic systems furnishes in turn a crucial link between this
escape rate formula and transport coefficients in nonequilibrium
steady states, such as diffusion coefficients, viscosities, conductivities
and chemical reaction rates \cite{GN,GD1,GD2,gaspard_chaos_1998,dorfman_introduction_1999,Kla06,Kla17},
substantiating the escape rate formalism that we already mentioned
above.

Along these lines, a long-standing fundamental open question is
whether there is any generalisation of this escape rate formalism to
weakly chaotic dynamical systems
\cite{korabel_deterministic_2004,Kla06,klages_weak_2013}. If so, it
should pave the way to express generalised transport coefficients
characterising anomalous transport in terms of corresponding
generalised chaos quantities assessing weak chaos
\cite{klages_weak_2013}. The simplest systems that can be studied
along these lines are time-discrete one-dimensional maps. Despite
their simplicity, they have been shown to nevertheless reproduce
fundamental chaotic and statistical-physical properties as exhibited
by much more complex, realistic systems
\cite{Schu,Ott,gaspard_chaos_1998,dorfman_introduction_1999,Voll02,Kla06}. An elementary example
reproducing strong chaos is the famous Bernoulli map (or: doubling map,
shift map, dyadic transformation) as discussed in textbooks
\cite{dorfman_introduction_1999,Kla17,Schu,Ott}. A paradigmatic generalisation, containing
the Bernoulli map as a special case, which models weakly chaotic
dynamics, is the Pomeau-Manneville map referred to above. It generates a mode
of dynamics that is called intermittent, as it alternates between laminar
phases, where a particle sticks to a marginally unstable fixed point,
and chaotic bursts, where a particle moves away seemingly randomly to
other parts in the phase space \cite{Schu,Ott,klages_weak_2013}. For
the closed system a generalised version of the Pesin identity could be
obtained
\cite{gaspard_sporadicity_1988,korabel_deterministic_2004,KoBa09,KoBa10,saa_pesin-type_2012,korabel_numerical_2013,klages_weak_2013,naze_number_2014}.
However, results for an open version of this simple weakly chaotic
model with escape are scarce in the literature. There are rigorous
mathematical results for the parameter regime of this map where it
still exhibits a positive Lyapunov exponent
\cite{Dahl99,FMS11,demers_escape_2016,DeTo17} but, to our knowledge,
not for the one where the Lyapunov exponent is zero.

Accordingly, the goal of our paper is to assess essential dynamical
systems properties of the open Pomeau-Manneville map, especially in
the parameter region where its dynamics is weakly chaotic, featuring a
zero Lyapunov exponent. Our article is structured as follows: in order
to set the scene by providing relevant background knowledge, in Sec.~2
we start with a brief review of the closed Pomeau-Manneville map and
its characteristic properties, which touches upon concepts of infinite
ergodic theory. In Sec.~3 we introduce the open Pomeau-Manneville map,
where we focus in particular on the spreading between nearby
trajectories that is assessed by the Lyapunov exponent. However, as
this exponent yields by definition a value of zero for subexponential
spreading, we refine our approach by instead looking at a generalisation,
which we call Lyapunov `stretching', which corresponds to the Birkhoff
sum over all contributions to the stretching. Its time evolution
yields important information about the irregularity of trajectories,
and eventually determines the Lyapunov exponent when obtained as a
time average. We calculate this key quantity numerically from simulations,
and analytically through a stochastic model,
and compare the results with each other. Surprisingly, we
find that the mean cumulative stretching approaches a constant value with
time, instead of growing without bound as in strongly chaotic dynamical
systems, which, counterintuitively, shows that in the long-time limit there is
no dynamical instability generating irregular spreading of
trajectories any more. In Sec.~4 we compute the dynamical systems entropy
production in the open map and comment on the difficulty to obtain
reliable, conclusive results. In this section we also assess in detail
the structure of the fractal repeller emerging in the open system. In
Sec.~5 we extend our methods to the full range of parameters,
including those exhibiting positive Lyapunov exponent as studied previously,
where we confirm analogous results. We conclude in Sec.~6 by commenting about
the question of a generalised escape rate formula for this
system. Given our findings, the existence of such a formula is
impossible, in line with previous mathematical results
\cite{demers_escape_2016}. We discuss the consequences of this
conclusion for cross-linking anomalous transport coefficients to weak
chaos quantities by a potentially generalised escape rate approach for
weakly chaotic dynamical systems. A very brief account of the main
results of this work has been given in the conference proceedings
Ref.~\cite{BrKl24}.

\section{Background}

\begin{figure}
\centering
\includegraphics[width=0.55\linewidth]{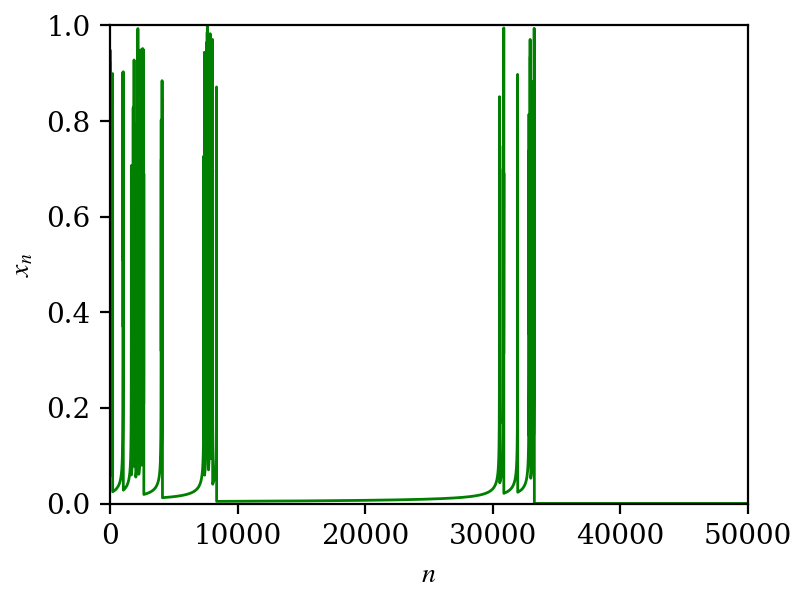}
\includegraphics[width=0.42\linewidth]{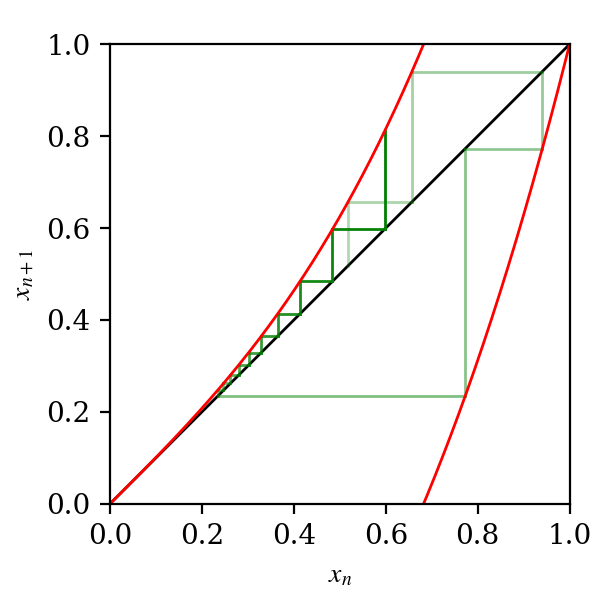}
\caption{{Left: A typical trajectory of the closed Pomeau-Manneville map as a function of $n$, displaying the periods of chaotic and laminar motions which give it the descriptor `intermittent'. Right: The closed PM map (red), with a sample trajectory shown as a cobweb plot (green). The line $x_{n+1}=x_n$ is shown for comparison (black).}}
\label{fig:newfig}
\end{figure}

\subsection{Pomeau-Manneville map}

In this paper we consider a variation on the Pomeau-Manneville map,
\begin{equation}
M(x) = x + ax^z \quad \textrm{(mod $1$)}
\end{equation}
for $z>1$, $a>0$, generating dynamics as $x_{n+1} = M(x_n)$ for a given initial condition $x_0$. In the closed map, we typically take $a=1$ so that $M(1^-)=1$; an example is shown in Fig.~\ref{fig:newfig}. This map was first described in \cite{manneville_intermittency_1979, manneville_intermittency_1980, pomeau_intermittent_1980} as a representation of the intermittent dynamics observed in Lorenz systems in models of atmospheric convection currents \cite{lorenz_deterministic_1963}.
The map is characterised by its {marginally unstable} fixed point at $x=0$, with $M'(0)=1$, whose existence is responsible for the peculiar dynamics of the map, which are characterised by contrasting periods of chaos (when $x_n$ is away from the fixed point) and slow, laminar motion (when $x_n$ is close to the fixed point), which define the dynamics as `sporadic' or `intermittent' \cite{gaspard_sporadicity_1988, geisel_anomalous_1984, geisel_accelerated_1985, zumofen_scale-invariant_1993, zumofen_power_1993}, {see Fig.~\ref{fig:newfig}}.

Near the fixed point, the laminar dynamics can be continuously approximated by \cite{korabel_fractal_2007, pegler_anomalous_2017, BSPKK25}
\begin{equation} \label{eq:dif-eq}
\dv{x}{t} = ax^z
\end{equation}
from which we may derive a characteristic time scale for the duration spent in the laminar phase by a particle {at position} $x_n$: if $x_n$ is injected uniformly into the region around the fixed point, the time until its escape is given by the density
\begin{equation} \label{eq:wt}
w(t) = \frac{\gamma b^\gamma}{(b+t)^{\gamma+1}}, \quad t>0
\end{equation}
for $\gamma := \frac{1}{z-1}$ and $b := \gamma/a$
\cite{geisel_anomalous_1984, zumofen_scale-invariant_1993,
  zumofen_power_1993, KCKSG06, korabel_fractal_2007, BSPKK25}. This
distribution, a power law, can be named as Pareto (Type II), Lomax,
or $q$-exponential, and has
diverging moments $\int_0^\infty t^k w(t) \dd{t}$ for $k\geq
\gamma$. Importantly, for $\gamma\leq 2$ (corresponding to map
parameters $z>\frac32$), $w(t)$ has no finite variance, and for
$\gamma\leq 1$ (resp.\ $z\geq 2$), $w(t)$ has no finite mean. {This
  separates the dynamics into three main dynamical regimes
  \cite{gaspard_sporadicity_1988}, shown in Table~\ref{tab:table1},
  which we must consider separately. We also show in
  Table~\ref{tab:table1} the special case $z=1$ ($\gamma=\infty$), for
  which the map is piecewise linear and uniformly hyperbolic, and for
  which the formula \eqref{eq:wt} no longer holds.}

\begin{table}
\centering
\small
\begin{NiceTabular}[hvlines]{cccccc}
 & \Block{}{(uniformly)\\ hyperbolic?\\ $\abs{M'(x)}>1$}
 & \Block{}{strongly\\ chaotic?\\ $\lambda>0$}
 & \Block{}{normalisable\\ invariant\\ density?}
 & \Block{}{recurrence\\ times}
 & \Block{}{growth of\\ dynamical\\ quantities} \\
\Block{}{$z=1$\\ (piecewise\\ linear)}
 & \Block[fill=green!25]{}{\cmark}
 & \Block[fill=green!25]{}{\cmark}
 & \Block[fill=green!25]{}{\cmark}
 & \Block[fill=green!25]{}{all moments\\ finite}
 & \Block[fill=green!25]{}{exactly linear} \\
$1<z<\frac32$
 & \Block[fill=red!25]{}{\xmark}
 & \Block[fill=green!25]{}{\cmark}
 & \Block[fill=green!25]{}{\cmark}
 & \Block[fill=green!25]{}{finite mean\\ and variance}
 & \Block[fill=green!25]{}{linear\\ (with normal\\ fluctuations)} \\
$\frac32<z<2$
 & \Block[fill=red!25]{}{\xmark}
 & \Block[fill=green!25]{}{\cmark}
 & \Block[fill=green!25]{}{\cmark}
 & \Block[fill=yellow!25]{}{finite mean,\\ diverging\\ variance}
 & \Block[fill=yellow!25]{}{linear\\ (with Lévy\\ fluctuations)} \\
$z>2$
 & \Block[fill=red!25]{}{\xmark}
 & \Block[fill=red!25]{}{\xmark}
 & \Block[fill=red!25]{}{\xmark}
 & \Block[fill=red!25]{}{diverging mean\\ and variance}
 & \Block[fill=red!25]{}{non-linear\\ (and Lévy\\ distributed)}
\end{NiceTabular}
\caption{{A table showing the key differences between the three main dynamical regimes of the closed Pomeau-Manneville map, and the piecewise linear special case $z=1$, in terms of their important dynamical properties. In open systems the piecewise linear case is well understood; in this paper we focus mainly on the case $z>2$, but we remark upon the other two regimes in Sec.~\ref{sec:otherz}.}}
\label{tab:table1}
\end{table}

The diverging moments of $w(t)$ cause quite some problems: for $z\geq 2$, the map has no normalisable invariant measure, but rather preserves an \emph{infinite invariant measure}
\cite{gaspard_sporadicity_1988, aaronson_introduction_1997, akimoto_subexponential_2010, thaler_transformations_1983, klages_weak_2013}
which is no longer normalisable. As a result, there is no convergence to a non-equilibrium steady state represented by a true, {normalisable} invariant {probability} density, but rather a continued evolution of the map's density towards the infinite invariant density \cite{korabel_numerical_2013, akimoto_aging_2013}. The non-existence of steady states implies that the system's dynamics are susceptible to vary with time, a phenomenon known as \emph{dynamical aging} \cite{bouchaud_weak_1992, barkai_aging_2003, akimoto_aging_2013}, observed in systems from the stochastic \cite{barkai_aging_2003-1, magdziarz_aging_2017} to the physical \cite{monthus_models_1996, brokmann_statistical_2003, barkai_aging_2004} and financial \cite{cherstvy_time_2017}.

As a result, the map is no longer ergodic \cite{zweimuller_ergodic_1998, zweimuller_ergodic_2000} in the sense that it no longer obeys the Birkhoff formula \cite{birkhoff_proof_1931}
\begin{equation}
\frac{1}{n} \sum_{k=0}^{n-1} f(x_k) \to \int f \dd{\mu^*} \quad (n\to\infty),
\end{equation}
for $\mu^*$ the normalised invariant measure, for Lebesgue-typical $x_0$. The failure of this relation is known to physicists as \emph{weak ergodicity breaking} \cite{klages_weak_2013, bouchaud_weak_1992, bel_weak_2005, bel_weak_2006}. Instead, time-averaged variables remain a function of the initial condition $x_0$ (ie.\ `random'), and we obtain only convergence in distribution, of the form
\begin{equation}
\frac{1}{a_n} \sum_{k=0}^{n-1} f(x_k) \xrightarrow{\mathrm{d}} \xi_\alpha \int f \dd{\mu^*}
\end{equation}
for some sequence $(a_n)$, and some random variable $\xi_\alpha$ which is Mittag-Leffler distributed with index $\alpha \in (0,1)$ and $\Exp[\xi_\alpha]=1$. This result is known as the Aaronson-Darling-Kac (ADK) theorem \cite{darling_occupation_1957, aaronson_asymptotic_1981, aaronson_introduction_1997, thaler_distributional_2006, akimoto_subexponential_2010, pires_lyapunov_2011, akimoto_aging_2013, klages_weak_2013}.
For the Pomeau-Manneville map, $(a_n)$ grows {proportionally to} $n^\gamma$ \cite{zweimuller_ergodic_2000}.

In this paper we interest ourselves in two quantities: the Lyapunov `stretching', which we define as a cumulative generalisation of the Lyapunov exponent, and the generation of entropy. The \emph{Lyapunov stretching} we define as
\begin{equation} \label{eq:stretching-def}
\Lambda_n := \sum_{k=0}^{n-1} \ln \abs{M'(x_k)}
\end{equation}
from which the traditional Lyapunov exponent is recovered by
\begin{equation}
\lambda = \lim_{n\to\infty} \frac{1}{n} \Lambda_n.
\end{equation}
We make this definition in order to generalise the Lyapunov exponent to weakly chaotic systems, namely those for which $\lambda = 0$; in such systems, $\Lambda_n$ increases sublinearly \cite{korabel_deterministic_2004,klages_weak_2013}. In the Pomeau-Manneville map, for $z<2$ (ie.\ $\gamma>1$) we have $\lambda>0$ (and thus $\Lambda_n \simeq \lambda n$), while for $z>2$ (ie.\ $\gamma<1$) we have $\langle \Lambda_n \rangle \sim n^\gamma$, which was conjectured in \cite{gaspard_sporadicity_1988}, shown numerically in \cite{korabel_deterministic_2004,KoBa09,KoBa10,korabel_numerical_2013} and proven in \cite{naze_number_2014, saa_pesin-type_2012,klages_weak_2013}.

This state of affairs is summarised in the first row of Table~\ref{tab:table2}, indicating the rate of growth of $\Lambda_n$ in the different parameter ranges of the map. Below we {outline} the method used by \cite{gaspard_sporadicity_1988} to estimate $\Lambda_n$ applied to the closed map. We will later adapt this method to the open system, which will form the main part of our results.

\begin{table}
\centering
\small
\begin{NiceTabular}[hvlines]{cccc}
\Block[fill=gray!25]{}{\phantom{m}\\ \phantom{m}\\ \phantom{m}\\ \phantom{m}}
 & \Block[fill=gray!25]{}{\textbf{strongly chaotic}\\
  and \textbf{unif.\ hyperbolic}\\
  (e.g.\ piecewise linear)}
 & \Block[fill=gray!25]{}{\textbf{strongly chaotic}\\
  and \textbf{non-hyperbolic}\\
  (e.g.\ PM, $1<z<2$)}
 & \Block[fill=gray!25]{}{\textbf{weakly chaotic}\\
  (e.g.\ PM, $z>2$)} \\
\Block[fill=gray!25]{}{\phantom{m}\\ \textbf{closed system}\\ (no escape)\\ \phantom{m}}
 & \Block{1-2}{\large $\Lambda_n \sim n$}
 && \Block{}{\large $\Lambda_n \sim n^\gamma$} \\
\Block[fill=gray!25]{}{\phantom{m}\\ \textbf{open system}\\ (escape)\\ \phantom{m}}
 & \Block{}{\textbf{exponential escape} \vspace*{2pt} \\
  {\large $\langle \Lambda_n \rangle \sim n$}\\
  {\footnotesize on the fractal repeller}}
 & \Block{1-2}{\textbf{algebraic escape} \vspace*{5pt} \\
  {\large $\langle \Lambda_n \rangle \sim\ ???$}}
\end{NiceTabular}
\caption{A table showing, in very broad terms, the known state of affairs for the dynamical and escape properties of open and closed hyperbolic, non-hyperbolic and weakly chaotic dynamical systems, considering the Pomeau-Manneville map as our frame of reference. Our work focuses mainly on the dynamical quantities of the lower-right quadrant, see \cite{demers_escape_2016} for escape properties and Sec.~\ref{sec:otherz} for a discussion of dynamics in the central $1<z<2$ regime.}
\label{tab:table2}
\end{table}

\subsection{Gaspard-Wang theory}

\begin{figure}
\centering
\includegraphics[width=0.4\linewidth]{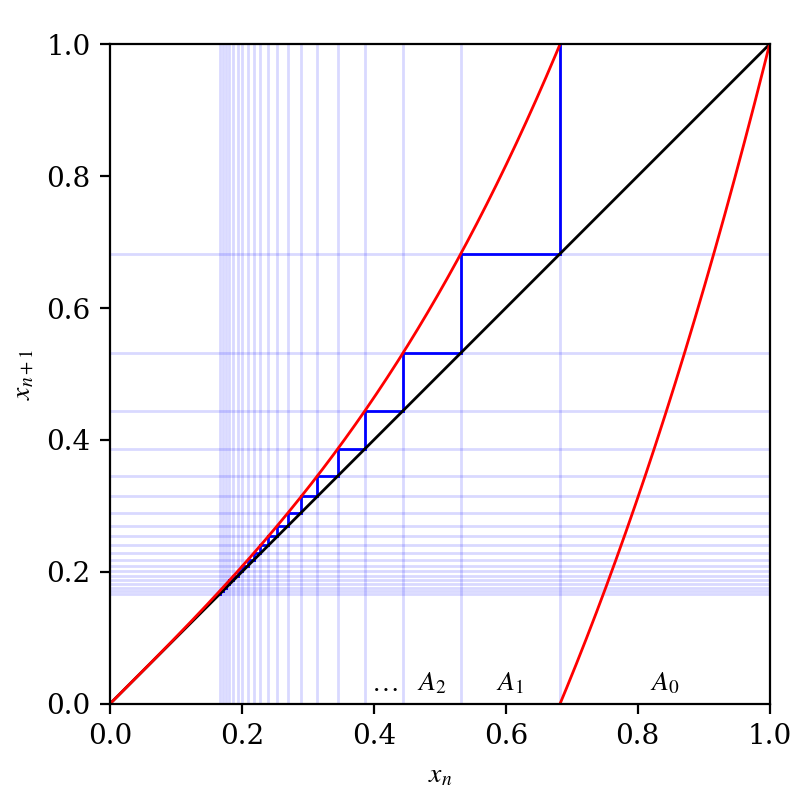}
\includegraphics[width=0.4\linewidth]{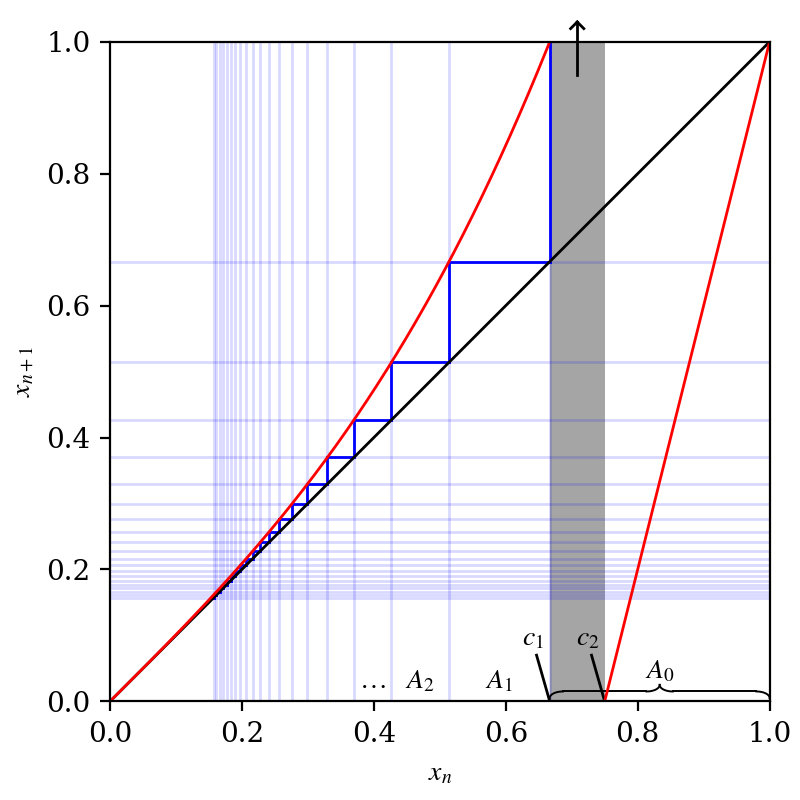}
\caption{Left: The closed Pomeau-Manneville map shown in red for $z=3$, $a=1$, with its corresponding Gaspard-Wang partition shown in blue. The diagonal $x_{n+1}=x_n$ is shown in black. Right: The open Pomeau-Manneville map for $z=3$, $c_1=\frac23$, $c_2=\frac34$, with its corresponding adapted Gaspard-Wang partition, with the escape region $(c_1,c_2)$ shown in grey.}
\label{fig:map}
\end{figure}

In \cite{gaspard_sporadicity_1988}, the authors defined a countable partition of the unit interval $\{A_0,A_1,A_2,\dots\}$ defined by preimages of the map's branch cut, shown in Fig.~\ref{fig:map}. This generates a chain
\begin{equation} \label{eq:c-dyn}
\dots \to A_2 \to A_1 \to A_0 \to \begin{cases} A_0,\\ A_1,\\ A_2,\\ \dots \end{cases}
\end{equation}
whereupon each interval of the partition maps approximately uniformly onto the next, and $A_0$ maps bijectively onto the entire unit interval. If the map branch in each interval is linearised, as in \cite{gaspard_sporadicity_1988, Wang89a}, then the map preserves a uniform density within each interval. These dynamics define a countable Markov process where the transition probabilities at the final step are determined by the relative size of the intervals,
\begin{equation} \label{eq:c-trans}
\Prb[A_0 \to A_k] \sim (k+1)^{-\gamma-1}.
\end{equation}
In \cite{gaspard_sporadicity_1988} the authors investigated the recurrence properties of $A_0$, namely, the number of visits to $A_0$ in $n$ timesteps of the process, denoted by $N_n$. The calculations are simple, emerging as a special case of results in \cite{feller_fluctuation_1949}, and we have explicitly included them for the case $0<\gamma<1$ in Sec.~1.1 of the supplementary material, since we will use them later. The following results are obtained: if $\mu$ and $\sigma$ are the mean and standard deviation of the individual recurrence time in \eqref{eq:c-dyn} (which is related to \eqref{eq:wt}), where they exist, then
\begin{itemize}
\item if $\gamma>2$,
\begin{equation}
\Prb \left[N_n \geq \frac{n}{\mu} - x \frac{\sigma n^{1/2}}{\mu^{3/2}} \right] \to \Phi(x)
\end{equation}
\item if $1<\gamma<2$,
\begin{equation}
\Prb \left[ N_n \geq \frac{n}{\mu} - x \left( \frac{nA}{\mu^{1+\gamma}} \right)^{1/\gamma} \right] \to G_\gamma(x)
\end{equation}
\item if $0<\gamma<1$,
\begin{equation}
\Prb \left[ N_n \geq \frac{n^\gamma}{Ax^\gamma} \right] \to G_\gamma(x)
\end{equation}
\end{itemize}
for $\Phi(x)$ the cumulative distribution function of a standard normal random variable,
\begin{equation}
\Phi(x) = \frac12 \left[ 1 + \erf(x/\sqrt{2}) \right]
\end{equation}
and $G_\gamma(x)$ the cumulative distribution function of a Lévy distribution whose characteristic function is given by
\begin{equation}
\psi_\gamma(s) = \exp\left\{ -\abs{s}^\gamma \Gamma(1-\gamma) \left[ \cos(\frac{\pi\gamma}{2}) - \im \sin(\frac{\pi\gamma}{2}) \frac{s}{\abs{s}} \right] \right\}.
\end{equation}
In general, a closed form expression for $G_\gamma(x)$ is not known, with the exception of the case $\gamma=\frac12$ \cite{feller_introduction_1968, klages_anomalous_2008}, where
\begin{equation}
G_{1/2}(x) = 2 \left\{ 1 - \Phi \left[ \left( \frac{\pi}{2x} \right)^{1/2} \right] \right\}
\end{equation}
with density function
\begin{equation}
p(x) = \frac12 x^{-3/2} \exp(-\frac{\pi}{4x}).
\end{equation}
The expected value $\Exp[N_n]$ is shown to increase linearly with $n$, namely $\Exp[N_n] \simeq \frac{n}{\mu}$, in the cases covering $\gamma>1$, and algebraically in the case $\gamma<1$, with $\Exp[N_n]\sim n^\gamma$. This forms a central part of Gaspard and Wang's thesis, which is that both the Lyapunov stretching $\Lambda_n$ and the algorithmic complexity, which relates to the Kolmogorov-Sinai entropy, are asymptotically proportional to $N_n$, since $A_0$ and its neighbouring regions of phase space are the dominant contributors to these quantities.

\section{Main results}

\subsection{Open Pomeau-Manneville map}

In this paper we consider an open variant of the Pomeau-Manneville map, {also related to a map of Liverani et~al.\ \cite{liverani_probabilistic_1999},} which we define by
\begin{equation} \label{eq:open-map}
M(x) = \begin{cases}
x + ax^z, &0\leq x\leq c_1, \\
\frac{x-c_2}{1-c_2}, &c_2 \leq x\leq 1,
\end{cases}
\end{equation}
with $a := (1-c_1)c_1^{-z}$ defined so that $M(c_1)=1$, ie.\ so that both branches of the map span the interval $[0,1]$. The interval $c_1 < x < c_2$ defines the `escape region', and trajectories which fall into this region are disregarded and considered to be killed. By this construction, almost all trajectories (with respect to, say, the Lebesgue measure on $x_0$) will eventually leave the interval, with the proportion of trajectories remaining alive at time $n$, {which we call survival probability,} being asymptotically proportional to $n^{-\gamma}$ \cite{demers_escape_2016, korabel_deterministic_2004}. It is not difficult to observe that the support of those trajectories which are not killed -- the `fractal repeller' of the system \cite{kadanoff_escape_1984, kantz_repellers_1985, bohr_entropy_1987} -- resembles a deformed Cantor set, albeit the deformity is highly non-trivial.
For a trajectory with initial condition $x_0$ we denote the time at which the trajectory escapes by $t_\mathrm{esc}(x_0)$; if $x_0$ lies on the fractal repeller then $t_\mathrm{esc}(x_0)=\infty$.
In this paper, we take $(c_1, c_2) = (\frac23, \frac34)$, although any suitable values would suffice, {and from simulations} we do not believe these values to be in any way special. We also interest ourselves mainly in the dynamical regime where $z\geq 2$, and {for analytic convenience, our calculations are performed in the case} $z=3$ (corresponding to $\gamma=\frac12$), {exemplifying this regime.}

The current state of knowledge concerning the dynamics on the open system are thus summarised in the second row of Table~\ref{tab:table2}. This demonstrates the exponential escape and positive Lyapunov exponent that the piecewise linear open map is known to have, as well as the results from \cite{korabel_deterministic_2004, demers_escape_2016} concerning the escape rate in the non-linear ($z>1$) system. Ref.~\cite{demers_escape_2016} proves this result for the cases $z<2$ (cases for which a finite invariant density exists in the closed system), while \cite{korabel_deterministic_2004} gave numerical evidence supporting this in all areas, which we verified numerically. Ref.~\cite{demers_escape_2016} also showed that the evolving density in the open map conditioned on survival converges to the delta function at the marginal fixed point, $\delta(0)$; another fact which we also numerically confirm for all parameter ranges.

In this paper we are interested in identifying the Lyapunov stretching \eqref{eq:stretching-def} in the open Pomeau-Manneville map. In order to deal with the problem of trajectories escaping, we must condition our ensemble averages on the survival time of the trajectory. Therefore we define
\begin{equation}
\langle \Lambda_n \rangle_t := \langle \Lambda_n(x_0) \,|\, t_\mathrm{esc}(x_0) > t \rangle
\end{equation}
as the Lyapunov stretching at time $n$ {(interpreted as the physical system time)} conditional on the process surviving up to time at least $t$ {(interpreted as a measurement time)}, for $0\leq n\leq t$. Further, we are interested in the behaviour of this quantity on trajectories on the fractal repeller, ie.\ the set of trajectories which never escape. This is of interest to us because of its significance in the escape rate formalism, see Sec.~\ref{sec:pesin}. {This therefore motivates the reasoning that studying $\langle \Lambda_n \rangle_t$} as we increase {the measurement time} $t\to\infty$ will give us some indication of the dynamics on the repeller, as the repeller is by definition the limit set of trajectories surviving to $t=\infty$; formally, $\bigcap_{t=0}^{\infty} \{ x_0 \,|\, t_\mathrm{esc}(x_0) > t \}$. Naturally, for any fixed $t$, $\langle \Lambda_n \rangle_t$ is an increasing function of {the physical time} $n$, with $\langle \Lambda_0 \rangle_t = 0$, {since $\Lambda_n$ is cumulative}, but \emph{a priori} it is not clear that $\langle \Lambda_n \rangle_t$ should have any monotonic relation with $t$ {for fixed $n$, and in general we found some settings in which it does not.}

\begin{figure}
\centering
\includegraphics[width=0.45\linewidth]{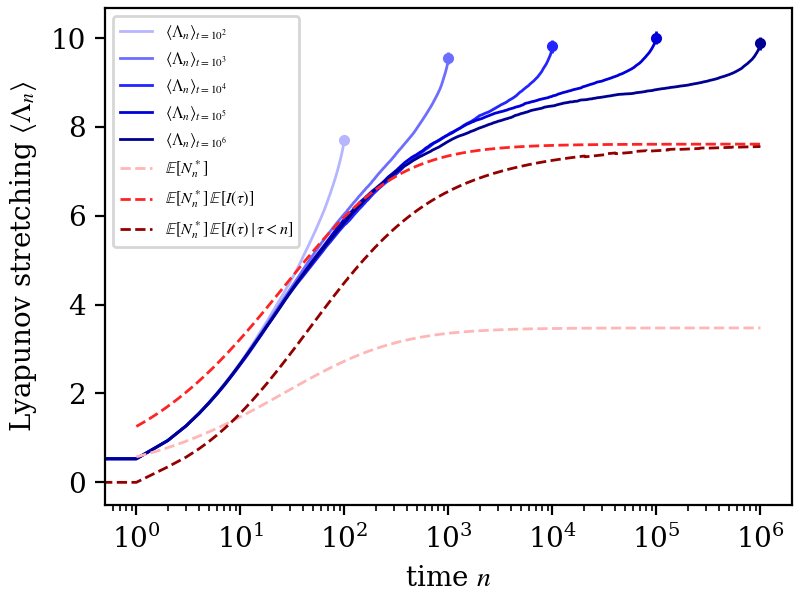}
\includegraphics[width=0.45\linewidth]{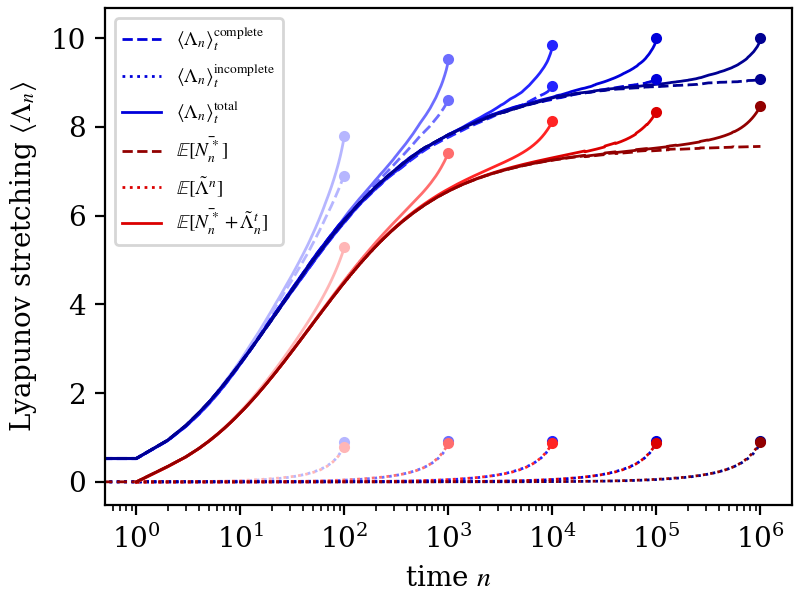}
\caption{Left: In shades of blue (solid), the cumulated Lyapunov stretching $\langle\Lambda_n\rangle_t$ computed from numerical simulations, as a function of $n$, for values of $t=10^i$, $i\in\{2,3,4,5,6\}$, $0\leq n\leq t$. Simulations are averaged over an ensemble of $N=10^4$ surviving trajectories. In shades of red (dashed), analytic estimates to $\langle \Lambda_n \rangle$ derived from the Gaspard-Wang theory, see legend and text for details. Right: In shades of blue (top), the cumulated Lyapunov stretching $\langle\Lambda_n\rangle_t$ incurred during complete (dashed) and incomplete (dotted, bottom) phases of each trajectory, with the total shown in solid lines. In red (below), estimates of the same, derived analytically from Gaspard-Wang theory, see text for details.}
\label{fig:results}
\end{figure}

Our initial numerical findings are shown in Fig.~\ref{fig:results}, for varying values of $t$ ranging between $t=10^2$ and $t=10^6$. We see that as $t$ increases, the stretching $\langle \Lambda_n \rangle_t$ for fixed $n$ appears to converge towards a distinctive functional form, which increases rapidly at first but later slows and flattens out. Further, it is unclear exactly how much this curve flattens out: whether it remains bounded as $(n,t)\to\infty$, or continues to increase at some rate, for example logarithmically. Finally, we see that, towards the end of each measurement period (ie.\ as $n\to t$), there is a systematic deviation from this functional form, in the form of a distinctive `tick' representing an increased amount of stretching. Our task in the following sections is to understand this behaviour in full via the use of an analytic stochastic model.

\subsection{Open Gaspard-Wang theory\label{sec:opengw}}

In this section we construct a stochastic model, {by generalising} the Gaspard-Wang model, to explain the flattening of the Lyapunov stretching, by demonstrating a corresponding flattening in $N_n$. We work in the case $z=3$, $\gamma=\frac12$, since this is the only case in the $\gamma<2$ regime for which we have an analytic expression for the appropriate corresponding Lévy law $G_{1/2}(x)$.

From Fig.~\ref{fig:map}, we construct a partition similarly to in \eqref{eq:c-dyn}, where we identify the union of the escape region and the right hand branch as $A_0 = (c_1, 1)$, and define $A_1, A_2, \dots$ as the preiterates of this region in the left hand branch. Then we can construct an identical symbolic dynamics on this new countable state space by
\begin{equation} \label{eq:o-dyn}
\dots \to A_2 \to A_1 \to A_0 \to \begin{cases} \mathrm{[esc]},\\ A_0,\\ A_1,\\ A_2,\\ \dots \end{cases}
\end{equation}
It is evident that the transition probabilities 
\begin{equation} \label{eq:o-trans}
\Prb[A_0 \to A_k] \sim (k+1)^{-\gamma-1}
\end{equation}
remain proportional to those in the closed case, but are rescaled by a factor of $(1-\varepsilon)$, for $\varepsilon := \frac{c_2 - c_1}{1-c_1}$ the probability of escape after having landed uniformly in $A_0$. Therefore, conditioning on trajectories who do not escape, the relative probabilities of reinjection to $A_k$ are equivalent to those in the closed case; except that \emph{trajectories with many reinjections are relatively \emph{de-weighted} in the ensemble, in exponential proportion to the number of reinjections they experience}: since each visit to $A_0$ comes with a $\varepsilon$-probability risk of escape, denoting by $N_n^*$ the number of visits to $A_0$ in $n$ timesteps in the open system,
\begin{equation}
p_{N_n^*}(x) \propto p_{N_n}(x) \cdot (1-\varepsilon)^x
 = p_{N_n}(x) \cdot \exp(-\nu x) \label{eq:bigf}
\end{equation}
for $\nu := -\ln(1-\varepsilon)$. In the supplementary material (Sec.~1.2) this is calculated for the case $\gamma=\frac12$, where it is revealed that $\Exp[N_n^*]$ converges to a positive constant as $n\to\infty$, with exponential fluctuations (ie.\ $p_{N_n^*}(x)$ converges to an exponential distribution). {In Fig.~\ref{fig:results} we plot $\Exp[N_n^*]$ in pale pink alongside the numerical results for the Lyapunov stretching $\langle \Lambda_n \rangle_t$.}

In order to compare this result with the Lyapunov stretching observed in simulations, we estimate the proportionality constant suggested in \cite{gaspard_sporadicity_1988} that relates $\langle \Lambda_n \rangle \simeq C \langle N_n \rangle$. In the supplementary material (Sec.~2) we roughly estimate the Lyapunov stretching incurred during a trajectory from $A_k$ through to $A_0$ asymptotically by
\begin{equation} \label{eq:i-tau}
I(k)\simeq (\gamma+1)\ln k + \Orb(1).
\end{equation}
Therefore we may take as an approximate constant $C := \Exp[I(k)]$ with $k$ power-law distributed according to \eqref{eq:c-trans}, equivalently \eqref{eq:wt} (shown in bright red in Fig.~\ref{fig:results}). However we may also take a more nuanced model, for example a time-dependent model $\langle \Lambda_n \rangle \simeq C_n \langle N_n \rangle$. For the remainder of this paper we use this approach with `constant' $C_n := \Exp[I(k)\,|\,k\leq n]$, conditioning simply on the fact that the duration of each completed excursion to the fixed point cannot exceed the total lifespan of the trajectory at that time {(see supplementary material, Sec.~2 for details).} We see in Fig.~\ref{fig:results} that this produces a curve, plotted in dark red, very similar in shape to the results from simulations, while differing quantitatively by some amount. This difference appears on the face of it to be approximately constant -- the curious reader is directed to Sec.~\ref{sec:repeller} and Fig.~\ref{fig:repeller}.

That the total stretching should flatten off so completely seems in some sense unphysical, and should be surprising, and one may cast doubt on whether this is truly what is observed in our simulations. In particular, we should seek a dynamical explanation of the `ticks' that are observed towards the end of each trajectory, as $n$ approaches $t$. Especially, the scale of these `ticks' should also remain bounded as $(n,t)\to\infty$ if the same is to be true of the stretching as a whole.

\begin{figure}
\centering
\includegraphics[width=0.98\linewidth]{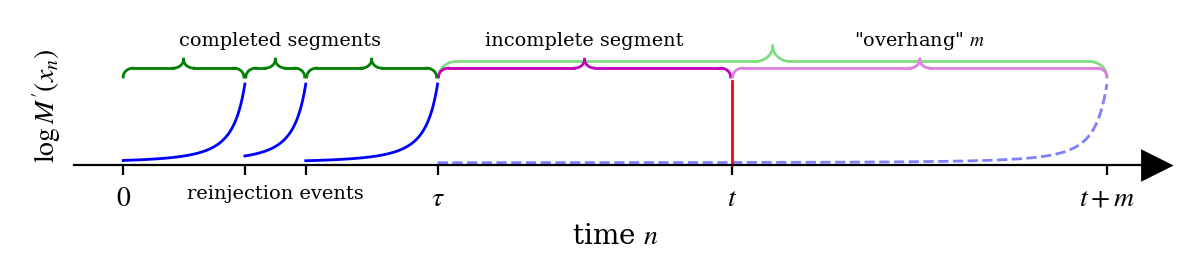}
\caption{A schematic diagram showing the separation of a typical trajectory into complete and incomplete segments. In blue: the instantaneous rate of stretching $\log M'(x_n)$. By estimating the `overhang' period $m$, we may estimate the contribution incurred during the incomplete phase.}
\label{fig:diagram}
\end{figure}

These may be explained by a more careful examination of the dynamics, which reveals a more subtle contribution which we show in Fig.~\ref{fig:diagram}. This depicts the instantaneous `stretching' $\log M'(x_n)$ as a function of time $n$ throughout several phases of a typical trajectory. Each visit to $A_0$ corresponds to the completion of a laminar motion from $A_k$ to $A_0$, which incurs some calculable amount of stretching, but there is also an `incomplete' motion which lasts from the time of the final visit to $A_0$, labelled $\tau$, until the end of the measurement time $t$. Since this segment of the trajectory never reaches $A_0$, its contribution to the stretching is not accounted for. Further, simulations reveal that typically for large $t$, this contribution tends to dominate the time interval, ie.\ $\tau \ll t$. Therefore it is important for us to understand the contribution from this `incomplete phase' of the trajectory.

In the supplementary material (Sec.~3) the contribution from the incomplete phase is {calculated via a stochastic model}, by estimating the time $m$ \emph{after} the end of measurement time, shown in the diagram (Fig.~\ref{fig:diagram}), at which an incomplete excursion would eventually reach $A_0$, and from this use \eqref{eq:i-tau} to calculate the contribution of the stretching within measurement time by
\begin{equation} \label{eq:i-tau-inc}
I_\mathrm{incomplete}([\tau, t]) = I_\mathrm{complete}([\tau, t+m]) - I_\mathrm{complete}([t, t+m])
\end{equation}
Here the `overhang' time $m$ is estimated via some delicate probability theory, see supplementary material (Sec.~3). By this approach we discover that the contribution of the incomplete segments corresponds exactly to the `ticks' observed in Fig.~\ref{fig:results}, and converges in the theory towards a standard functional form, as a function of $n/t$, whose height, or total contribution, is also asymptotically constant as $(n,t)\to\infty$. Adding these contributions, for varying $t$, to the raw $\Exp[\Lambda_n]$ gives a very good qualitative representation of the curves $\langle \Lambda_n \rangle_t$ observed in simulations, see Fig.~\ref{fig:results}. Further, by separating our simulation data into contributions from `completed' and `incomplete' phases (ie.\ stretching incurred before and after the final visit to $A_0$), we see evidence supporting our model: the complete contribution no longer displays ticks for large $(n,t)$, and instead displays a smooth, flat, convergent curve; while the incomplete contribution has a distinct repetitive functional form as predicted, which very strongly matches our model for the incomplete component.

\section{Generalised Pesin relation\label{sec:pesin}}

\subsection{Entropy}

\begin{figure}
\centering
\includegraphics[width=0.6\linewidth]{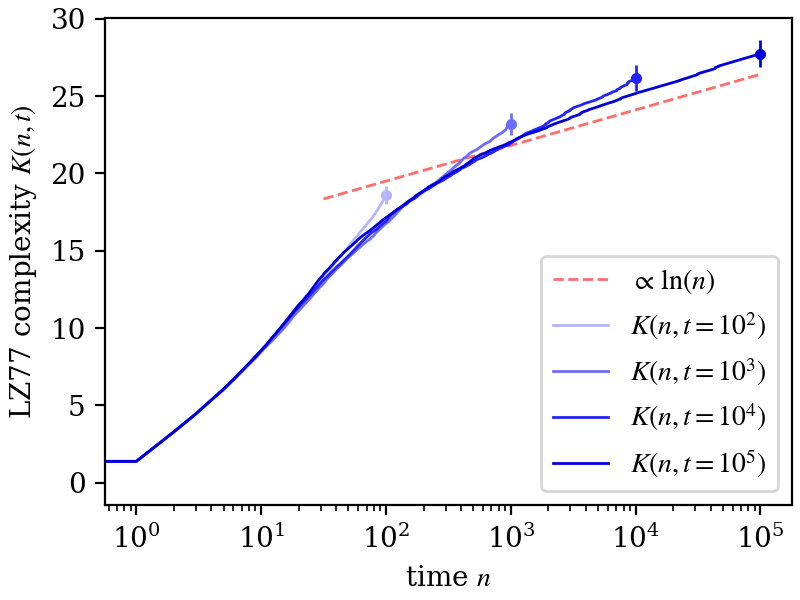}
\caption{In shades of blue: the LZ77 complexity $K(n,t)$ for $t=10^i$, $i\in\{2,3,4,5\}$, averaged over an ensemble of $N=10^3$ surviving trajectories. Error bars at the end of each line show the standard error about the sample mean. In faded red, dashed: the line proportional to $\ln(n)$ for comparison.}
\label{fig:entropy}
\end{figure}

Secondly, we concern ourselves with the generation of entropy in the open Pomeau-Manneville map. Specifically, the Pesin relation for strongly chaotic {($\lambda>0$)} closed systems, and the escape rate formalism for strongly chaotic open systems, concern themselves with the Kolmogorov-Sinai entropy $h_\mathrm{KS}$ over trajectories on the fractal repeller, defined by
\begin{equation}
h_\mathrm{KS} := \sup_\mathcal{A} \lim_{n\to\infty} \frac{1}{n} H_n, \qquad H_n := -\sum_{A_{n,i}\in\mathcal{A}_n} \mu(A_{n,i}) \ln \mu(A_{n,i}),
\end{equation}
for $\mathcal{A}_n$ the `refinement' of an initial partition $\mathcal{A}=\{A_i\}$ via backwards iteration \cite{gaspard_chaos_1998, dorfman_introduction_1999, badii_complexity_1997, castiglione_chaos_2008}.

Closely related to the Kolmogorov-Sinai entropy is the algorithmic complexity of Kolmogorov and Chaitin \cite{kolmogorov_combinatorial_1983, chaitin_algorithmic_1987, badii_complexity_1997}: given a partition $\{A_i\}$ of the system's state space, we record at each timestep $n$ the element of the partition into which $x_n$ falls, thus generating a sequence $\{w\}_n$ representing the symbolic dynamics of the process; then the \emph{algorithmic complexity} of the resulting sequence measures the minimum required length of a (binary) computer program $s(\{w\}_n)$ to reproduce it. For strongly chaotic systems, the sequence $w_n$ may appear effectively random, and thus $\len(s(\{w\}_n))\sim \len(\{w\}_n) = n$. For systems with periodic or highly regular dynamics, we may obtain $\len(s(\{w\}_n)) \ll n$. Here we define
\begin{equation} \label{eq:Kn-def}
K(n) = \langle \len(s(\{w\}_n)) \rangle \ln 2.
\end{equation}
Providing that the chosen partition satisfies certain requirements \cite{castiglione_chaos_2008, collet_iterated_1980}, we get that \cite{shannon_mathematical_1948, huffman_method_1952, alekseev_symbolic_1981, brudno_entropy_1983}
\begin{equation} \label{eq:Kn-hKS-equiv}
K(n) \geq H_n \quad \textrm{and} \quad \lim_{n\to\infty} \frac{1}{n} K(n) = h_\mathrm{KS}.
\end{equation}
For a strongly chaotic system (in one dimension) with $h_\mathrm{KS}>0$, the Pesin theorem relates the entropy to the Lyapunov exponent
\begin{equation}
\lambda = h_\mathrm{KS}
\end{equation}
and from \eqref{eq:Kn-hKS-equiv} we obtain $H_n \simeq K(n) \sim n$. For a periodic or constant sequence, we have $K(n) \sim \ln(n)$.

In order to estimate the algorithmic complexity, we tested the LZ77 \cite{lempel_complexity_1976} and LZ78 \cite{ziv_compression_1978} algorithms, which were designed as general purpose algorithms for text compression, and the CASToRe algorithm \cite{argenti_information_2002}, designed to be more sensitive than LZ78 for use on dynamical systems. For a partition we take the binary partition defined by the two branches of the map. In \cite{benci_information_2001, argenti_information_2002, benci_dynamical_2004, korabel_deterministic_2004}, each of these algorithms were used to estimate the entropy generation in some well known dynamical systems, including the closed Pomeau-Manneville map and others. The results were found to fit well with expectations, with $K(n) \sim n^\gamma$, which would indicate at a more general relation of the type
\begin{equation}
\langle \Lambda_n \rangle \sim H_n
\end{equation}
which would constitute a \emph{generalised Pesin relation}. This relation was later shown for closed Pomeau-Manneville type systems in \cite{naze_number_2014, saa_pesin-type_2012}.

We apply this to the surviving trajectories of the open system in the same way as we did to $\Lambda_n$, by defining $K(n,t)$ similarly to \eqref{eq:Kn-def}, where we condition the ensemble average in that equation on $t_\mathrm{esc}(x_0) \geq t$ as before. Our results {for the LZ77 algorithm} are shown in Fig.~\ref{fig:entropy}. However we do not find our results to be {conclusive}, for the following reasons: both the LZ78 and CASToRe algorithms have an in-built hard limit to their sensitivity, being unable to compress a binary string of length $n$ to anything less than $\Orb(n^{1/2})$ and $\Orb(\log_2 n)$ respectively, which our simulations run firmly up against (and therefore our results for these algorithms are not shown); and therefore we surmise $K(n) = \Orb(\log n)$ or less. Meanwhile, LZ77, while less computationally efficient, does not appear to have such a hard limit by its definition, and for this we also find $K(n) \sim \log n$. On the other hand, we know that the complexity of a periodic or constant sequence must also be at least $K(n) \sim \log n + \mathit{const.}$ \cite{benci_dynamical_2004, bonanno_computational_2002}. In practice we find our sample trajectories to indeed be very regular, as the measurement time is dominated by the phase of laminar motion in the left-hand branch (the `incomplete phase' described above), during which no `information' is added due to its regular dynamics. This, therefore, while {not fully constituting a positive} result, for us is nevertheless informative concerning the limitations of {using compression algorithms} as a proxy for entropy. {We consider our results to be consistent with a totally flattened entropy, matching the Lyapunov stretching, since we understand \emph{a priori} that this method would not be able to distinguish a total flattening from a logarithmic or sub-logarithmic growth. In this sense we offer here logarithmic growth as an upper bound to the `true' entropy $H_n$.}

\subsection{Analysis of the fractal repeller\label{sec:repeller}}

\begin{figure}
\centering
\includegraphics[width=0.5\linewidth]{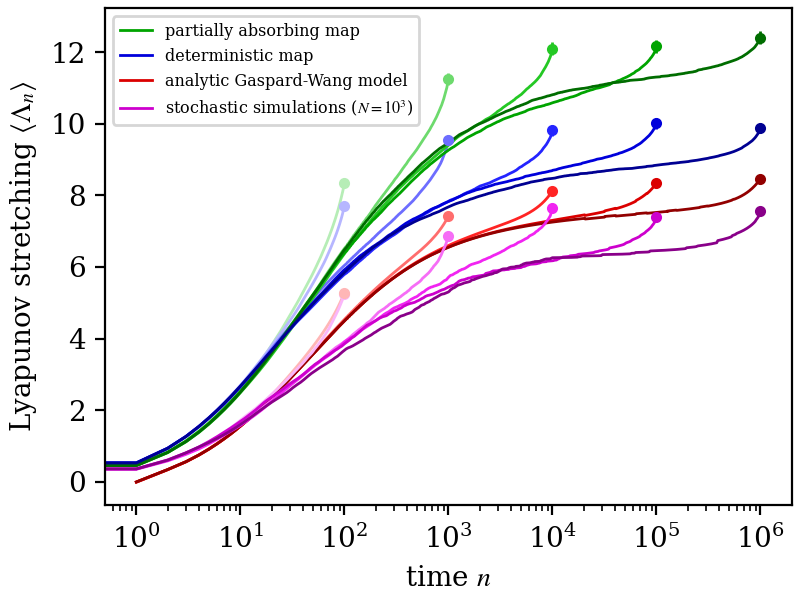}
\includegraphics[width=0.4\linewidth]{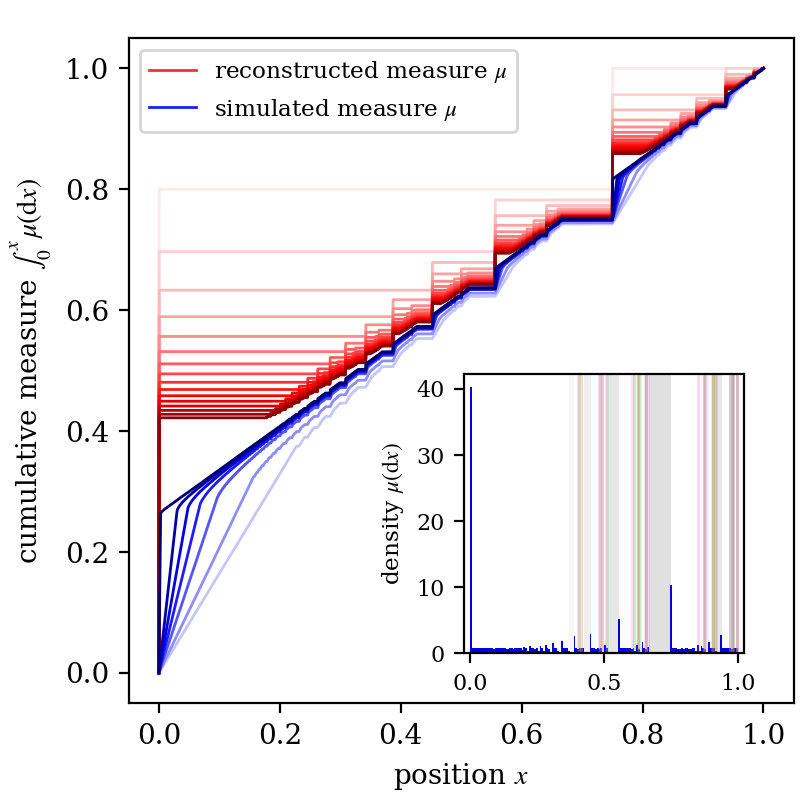}
\caption{Left: Lyapunov stretching $\langle \Lambda_n \rangle_t$ for $t=10^i$, $i\in\{2,3,4,5,6\}$ for, from top to bottom, the partially absorbing map variant (from simulations, in green), the deterministic map (from simulations, in blue), the analytic {open} Gaspard-Wang model (see Sec.~\ref{sec:opengw}, in red), and from stochastic simulations (see text, in magenta). Simulations for the deterministic and partially absorbing maps are averaged over an ensemble of $N=10^4$ surviving trajectories; the stochastic simulations are averaged over an ensemble of $N=10^3$ surviving trajectories. Right: Estimates of the cumulative measure on the fractal repeller of the deterministic open Pomeau-Manneville map, for $z=3$, taken via two limiting procedures: in blue, the cumulative measure of $N=10^6$ surviving initial conditions after $t$ iterations of the map, initialised from a uniform density, for values of $t$ between $t=10$ and $t=50,000$; in red: the reconstructed cumulative measure after $t$ iterations of the procedure described in the text, for values of $t$ between $t=1$ and $t=16$; in both cases, darker colours (closer to the centre of the plot) correspond to larger $t$; {inset: a histogram of the simulated measure, showing the approximate density function; grey vertical bars indicate the escape region and its preiterates; coloured vertical lines indicate prominent periodic points of varying orders and their preiterates; observe that the repeller is clustered on the fixed point $x=0$ and its preiterates.}}
\label{fig:repeller}
\end{figure}

Based on our discussion above, we now turn to analyse the structure of the fractal repeller of the open system. We would expect its structure to be extremely influential on the dynamics of the process, especially for longer measurement times, and in particular, we may be concerned that the open map restricted to the repeller (which is then a closed process) may still produce some non-zero amount of stretching, which is atypical of a standard trajectory and would therefore cast doubt on the suitability of a stochastic model.

This we approach in two ways: firstly, we compare the results from the deterministic map \eqref{eq:open-map} and the Gaspard-Wang calculations with two other systems. The first is a partially absorbing map, given by
\begin{equation}
M(x) = \begin{cases}
x + ax^z, &0\leq x\leq c_1, \\
\frac{x-c_1}{1-c_1}, &c_1 < x\leq 1,
\end{cases}
\end{equation}
{(taking $a$ and $c_1$ as before)} which is a closed map in which the escape region is unified with the right-hand branch, which instead becomes partially absorbing, with a particle instead being randomly ejected from the system with probability $\varepsilon := \frac{c_2 - c_1}{1-c_1}$ upon each visit to the right-hand branch. Thus, surviving trajectories undergo deterministic dynamics similar to in the ordinary map (and in the model), but the influence of the initial condition -- and therefore the fractal repeller -- on escape is suppressed.
The second is a purely stochastic simulation of the process defined by the open Gaspard-Wang model, based on the model process described in Fig.~\ref{fig:diagram}: we sequentially draw recurrence times from the time distribution \eqref{eq:wt} until their sum surpasses the prescribed measurement time $t$; however after each recurrence event we kill the process with some probability $\varepsilon$, after which we do not consider it. For each realisation of this process which survives until the measurement time is reached, we calculate the Lyapunov stretching based on the sampled recurrence times via \eqref{eq:i-tau} and \eqref{eq:i-tau-inc} (for complete and incomplete phases respectively). This therefore simulates precisely the stochastic setup defined by the open Gaspard-Wang model, and is unaffected by any potential {artefacts, deviations or subtleties} arising from the deterministic dynamics. In both cases we find results which qualitatively match those from the map and the model, see Fig.~\ref{fig:repeller}, but again differ quantitatively by some amount which appears asymptotically constant. In other words: introducing/removing aspects of the deterministic map (resp., stochastic aspects) to the system does not qualitatively change the observed generation of Lyapunov stretching.

Secondly, we look at the distribution of initial conditions which survive for very long times, as a proxy for the repeller itself. This is shown as a cumulative distribution in Fig.~\ref{fig:repeller}. We also plotted these as a fine histogram for $t=50,000$ (inset). In particular we looked for the prevalence of periodic and pre-periodic points, which could have a more prominent influence in the deterministic map that is not captured by the stochastic model. In strongly chaotic systems, periodic cycles often form a significant structure governing the dynamics \cite{artuso_recycling_1990, cvitanovic_chaos_2020}. We also looked at the same for the partially absorbing map (for which we would expect periodic cycles to be under-represented). However, we found no evidence of a special prevalence of periodic cycles in the ensemble; rather, we found only a dependence on pre-images of the fixed point at $x=0$.

In fact we hypothesise that this is the \emph{only} significant feature of the fractal repeller: in Fig.~\ref{fig:repeller} we show that we may reconstruct the distribution of surviving initial conditions by only considering scaled pre-iterates of the fixed point, via a procedure described in the supplementary material (Sec.~4), and we find that this converges quite well to the distribution obtained from simulations. In the reconstruction, the measure at each pre-image $x_{n-1}$ of some point $x_n$ is scaled in inverse accordance to the stretching incurred during the mapping from $x_{n-1}$ to $x_n$, ie.\
\begin{equation}
\mu(\dd{x_{n-1}}) \approx \frac{1}{\abs{M'(x_{n-1})}}\,\mu(\dd{x_n}).
\end{equation}
A more detailed description of the reconstruction of the repeller is given in the supplementary material (Sec.~4). This, we believe, goes some way to explaining the total suppression of stretching in the map, as only trajectories stuck close to pre-images of the marginal fixed point remain prominent in the system for long time. The backwards-iterates of the marginal fixed point form a skeleton for the dynamics of the open Pomeau-Manneville map in just the same way as periodic points form a skeleton for cycle expansions in hyperbolic maps \cite{artuso_recycling_1990, dettmann_cycle_1997, dettmann_computing_1998, artuso_cycle_2003, cvitanovic_chaos_2020}, except that while the separation of trajectories in the neighbourhood of periodic cycles has an exponential rate, trajectories near the marginal fixed point separate at a much `stickier' stretched exponential rate. This `collapse' of the repeller to the fixed point also explains the total suppression of the Lyapunov stretching for long times.

\section{Parameter variation\label{sec:otherz}}

\begin{figure}
\centering
\includegraphics[width=0.45\linewidth]{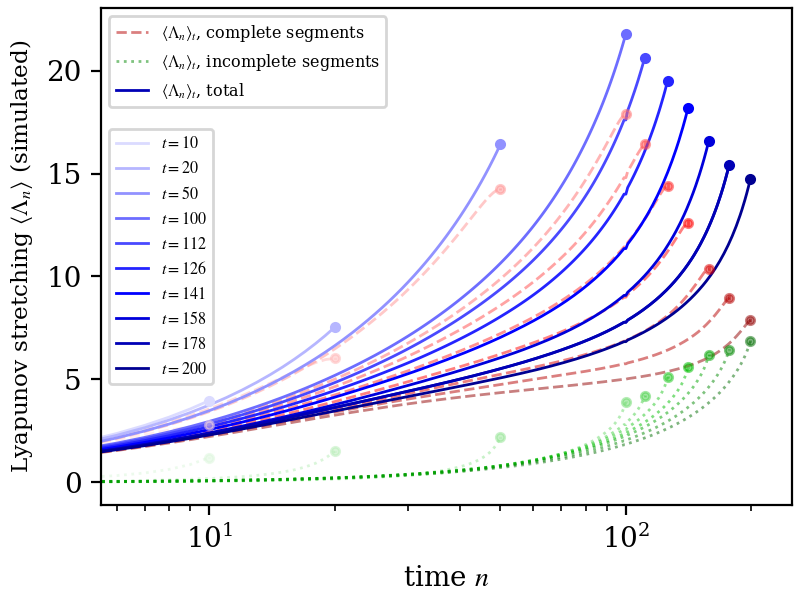}
\includegraphics[width=0.45\linewidth]{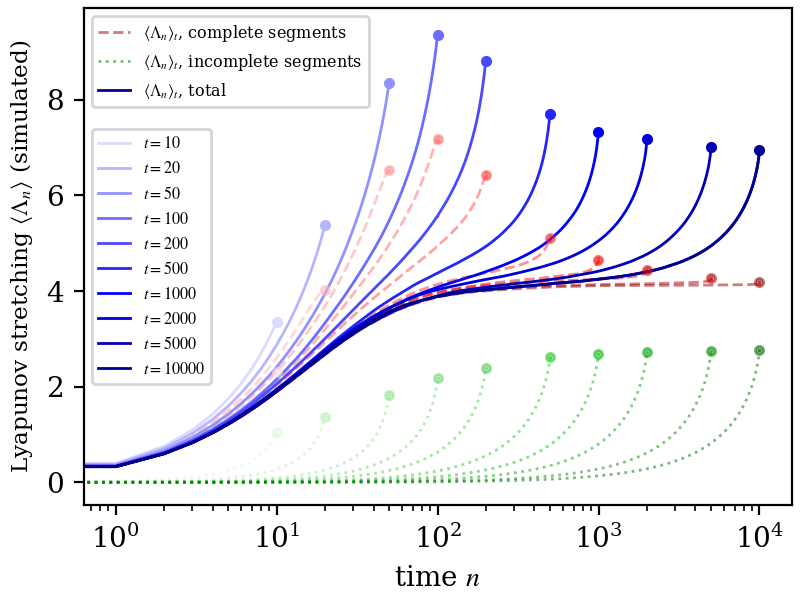}
\caption{Lyapunov stretching $\langle \Lambda_n \rangle_t$ from the purely stochastic simulated system described in the text, for $z=1.25$ (left) and $z=1.75$ (right), divided into the cumulated contributions from complete (red, dashed) and incomplete (green, dotted) phases (with the total in blue, solid lines), for values of $t$ ranging between $t=10$ and $t=200$ for $z=1.25$, and $t=10$ and $t=10^4$ for $z=1.75$, in both cases averaged over an ensemble of $N=10^4$ surviving trajectories.}
\label{fig:other-z}
\end{figure}

Finally, we repeat our analysis for a selection of different values of the parameter $z$. Since the closed Pomeau-Manneville map is known to pass through three distinct dynamical regimes, when $1<z<\frac32$ ($\gamma>2$, strongly chaotic with Gaussian fluctuations), $\frac32<z<2$ ($1<\gamma<2$, strongly chaotic with Lévy fluctuations), and $z>2$ ($0<\gamma<1$, weakly chaotic -- see Table~\ref{tab:table1}); we examine here results from simulations for selected $z$ in each of these intervals. Since it is only in the $z>2$ regime where the usual symptoms of weak chaos (the vanishing of the Lyapunov exponent, the infinite invariant measure, and weak ergodicity breaking) are observed, it is not \emph{a priori} clear to what extent the results we see in this case will be replicated in others.

In Fig.~\ref{fig:other-z} we plot the Lyapunov stretching from stochastically simulated systems in the cases $z=1.25$ and $z=1.75$.
Due to very high rates of escape in smaller $z$ regimes, obtaining sufficiently many surviving trajectories for long times in the deterministic map is unviable, and so we instead stochastically simulate the process, using the procedure described in Sec.~\ref{sec:repeller}. In Fig.~\ref{fig:repeller} we verify that this qualitatively matches to the results from the deterministic map in the case $z=3$.

We see in the $z=1.75$ case a distinct flattening very similar to $z=3$, even though in this case a positive Lyapunov exponent and a normalisable invariant density both exist in the closed system. In the $z=1.25$ case the situation is less clear, and gathering data for surviving trajectories for long times is harder due to significantly higher rates of escape (here we obtained data only up to $t=200$), but by analysing `complete' and `incomplete' segments of trajectories separately, we see both that the contribution from `complete' segments decays with large $(n,t)$ to a constant in a similar fashion to the other two cases, and that the contribution from the `incomplete' phases appears as though it may remain bounded (although on this latter point the numerical evidence is not conclusive). This is a good fit with results proven in \cite{demers_escape_2016} on the open Pomeau-Manneville map for $z<2$, which showed for a wide class of initial densities that the escape rate from such systems is always algebraic, and the evolving density of surviving trajectories always converges to a delta function at the fixed point, $\delta(0)$. Although not covered in that reference, this is also true for $z\geq 2$, as we have confirmed numerically. Thus the critical feature in the suppression of the dynamics may not be the loss of strong chaos or ergodicity at $z=2$, but rather the loss of hyperbolicity which occurs for all $z>1$, even in strongly chaotic and ergodic regimes. These results, in all, encourage us to fill in the remaining quadrant in Table~\ref{tab:table2} by claiming $\langle \Lambda_n \rangle \sim \Orb(1)$ with algebraic escape throughout the entire indicated region.

\section{Conclusion}

In summary, we have shown that by inserting a hole into an archetypal
intermittent and weakly chaotic map, the chaos inside the system, here
represented by the generation of Lyapunov stretching, a generalisation
of the Lyapunov exponent, is killed, in the sense that the average
total Lyapunov stretching remains bounded. The generation of this
quantity in both the closed and open systems is analytically explained
fully by a simple stochastic model, which is verified by
simulations. The correspondence between map and model, as well as the
influence of the map's fractal repeller, is tested extensively via the
use of a partially absorbing model, and the structure of the repeller
is shown to collapse completely onto pre-iterates of the map's
marginal fixed point, which we are able to fully reconstruct. The
generation of entropy was also examined, via the use of standard
algorithms to estimate dynamical complexity. While, given the limited
power of these existing algorithms, our results for the entropy
production were not entirely conclusive, we argued that overall they
are in line with our findings for the Lyapunov stretching. Finally, we
have presented numerical evidence that a similar total flattening of
the stretching is observed even for parameters which, when the system is
closed, are ergodic and strongly chaotic. We may remark that we have chosen
to construct our hole in a rather specific way, but we expect qualitatively
similar results to hold for a wide variety of hole placements, given broadly
applicable results in \cite{demers_escape_2016}, although in general a
dependence of dynamical quantities on the size and position of holes must be
noted \cite{bunimovich_where_2011, knight_dependence_2012}.

The collapse of the measure for all parameter values, one of the main results of this paper, has in our view profound implications. While the closed system exhibits a generalised Pesin identity
\cite{gaspard_sporadicity_1988,korabel_deterministic_2004,KoBa09,KoBa10,saa_pesin-type_2012,korabel_numerical_2013,klages_weak_2013,naze_number_2014},
our result implies that there can be no generalised escape rate formula
for the open Pomeau-Manneville map, cross-linking Lyapunov
stretching and metric entropy production on the fractal repeller to
escape from it. The total collapse of the measure instead implies a
profound instability of this weakly chaotic map with respect to
perturbation by drilling in a hole \cite{demers_escape_2016}. This
is in sharp contrast to uniformly hyperbolic, strongly chaotic systems
like the Bernoulli map, where there is no collapse of the measure when
opening up the system. In a way, the collapse may be understood as
resulting from an interplay between an essentially exponential escape
mechanism (in the sense of eq.~\eqref{eq:bigf}) and a subexponential mixing
generated by the map's internal dynamics.
Viewed in this way, in the Pomeau-Manneville map the `exponential'
escape depletes the system faster than it can mix internally, hence
there is no remaining non-trivial dynamics on the emerging fractal
repeller, and the whole measure collapses onto the fixed point and its
pre-images. This is of course only a very crude, qualitative picture
of the whole situation, but it seems to be in line with the scenario
analysed rigorously mathematically in Ref.~\cite{DeTo17}. A further important
consequence of the non-existence of a generalised escape rate formula
for the open Pomeau-Manneville map is that that there cannot be any
escape rate formalism relating anomalous transport coefficients to
weak chaos quantities for this dynamics, in contrast to what was
firmly established for strongly chaotic diffusive systems \cite{GN,GD1,GD2}.
As it stands, there is no simple way to `cure' this deficiency, and any
alternative to relate weak chaos quantities to anomalous transport
coefficients appears to be unclear at present.

One may question the generality of these results beyond the simple
framework of the Pomeau-Manneville map. For strongly chaotic dynamical
systems it was shown that the escape rate formalism holds for a
hierarchy of different dynamical systems, bridging the gap between
abstract mathematical models, like one-dimensional maps and
two-dimensional baker transformations, to physically more realistic
dynamics of particle billiards, like Lorentz gases
\cite{gaspard_chaos_1998,dorfman_introduction_1999,Voll02,Kla06}.
Along the same lines, it would be
important to check for the relevance of our result for more realistic
weakly chaotic dynamical systems, perhaps even for Hamiltonian
dynamical systems when their dynamics is governed by fractal
hierarchies of islands of stability in phase space
\cite{CrKe08,Vene09}. Systems exhibiting infinite invariant densities,
similar to the Pomeau-Manneville map, that can be studied
experimentally are in turn subrecoil laser-cooled atoms, which may
point in a direction to bring such phenomena to experimental reality
\cite{BRA21}. We may also remark that the theory we developed is based
on a Markov chain model, in the spirit of the approach by Gaspard and
Wang that employed a similar methodology
\cite{gaspard_sporadicity_1988}; but Markov chains belong to the
stochastic world, which in turn suggests a generality of our result
beyond dynamical systems theory, opening up a wider variety of further
modelling opportunities.

\ack
The authors would like to acknowledge helpful conversations with Stefano Galatolo. This research utilised Queen Mary's Apocrita HPC (High Performance Computing) facility, supported by QMUL Research-IT. \cite{butcher_apocrita_2017}


\section*{References}
\bibliographystyle{iopart-num}
{
\bibliography{bib_total, summ45}
}

%
%

\end{document}


\title[Supplementary material]{Supplementary material: Does an intermittent dynamical system remain (weakly) chaotic after drilling in a hole?}

\author{Samuel Brevitt$^1$ and Rainer Klages$^{1,2}$}

\address{$^1$ Centre for Complex Systems, School of Mathematical Sciences, Queen Mary University of London, Mile End Road, London E1~4NS, UK}
\address{$^2$ London Mathematical Laboratory, 8 Margravine Gardens, London W6~8RH, UK}

\ead{s.brevitt@qmul.ac.uk}



\section{Gaspard-Wang theory\label{sec:gw}}

The following calculations are based on results in \cite{gaspard_sporadicity_1988, feller_fluctuation_1949}. In Sec.~\ref{sec:cl-gw} we reproduce these results for the closed model, and in Sec.~\ref{sec:op-gw} generalise to the open model.

\subsection{Closed Gaspard-Wang model\label{sec:cl-gw}}

Firstly, we remark that by construction, a particle injected into the region $A_k$ will take exactly $k$ timesteps to return to $A_0$. Therefore, if
\begin{equation}
\Prb[ A_0 \to A_k ] \sim (k+1)^{-\gamma-1}
\end{equation}
asymptotically for $k\to\infty$ (we assume no escape in the closed system), and if $X$ is the time between visits to $A_0$, then $X$ has cumulative distribution function (CDF)
\begin{equation}
F_X(t) \simeq 1 - At^{-\gamma}
\end{equation}
asymptotically as $t\to\infty$, for some choice of $A>0$. Let $S_k = X_1 + \dots + X_k$, with $X_1$, $\dots$, $X_k$ independent, and let $N_n$ denote the number of reinjections (visits to $A_0$) in $n$ timesteps. It is not difficult to see that $S_k$ and $N_n$ are related by the formula \cite{feller_fluctuation_1949}
\begin{equation}
\Prb[ N_n \geq k ] = \Prb[ S_k \leq n ].
\end{equation}
For large $n,k$, we can obtain distributions for these quantities by some known results in probability theory: if $0<\gamma<1$, we have \cite{feller_fluctuation_1949, gaspard_sporadicity_1988}
\begin{equation} \label{eq:eq1}
\Prb\left[ N_n \geq \frac{n^\gamma}{Ax^\gamma} \right] \to G_\gamma(x)
\end{equation}
as $n\to\infty$, where $G_\gamma(x)$ is the CDF of a Lévy stable law
\cite{feller_fluctuation_1949, feller_introduction_1968, penson_exact_2010, saa_alternative_2011} with characteristic function
\begin{equation}
\psi_\gamma(s) = \exp\left\{ -\abs{s}^\gamma \Gamma(1-\gamma) \left[ \cos(\frac{\pi\gamma}{2}) + \im\sin(\frac{\pi\gamma}{2})\frac{s}{\abs{s}} \right] \right\}.
\end{equation}

If $Q_n$ is a random variable with CDF $G_\gamma(x)$, then equation \eqref{eq:eq1} is asymptotically equivalent to
\begin{equation}
\Prb\left[ N_n \geq \frac{n^\gamma}{Ax^\gamma} \right] = \Prb\left[ Q_n \leq x \right]
\end{equation}
which we may transform to
\begin{align}
\Prb\left[ n^{-1}(AN_n)^{1/\gamma} \geq \frac{1}{x} \right] &= \Prb\left[ Q_n \leq x \right] \nonumber \\
\implies \Prb\left[ n(AN_n)^{-1/\gamma} \leq x \right] &= \Prb\left[ Q_n \leq x \right]
\end{align}
and thus we can write
\begin{equation} \label{eq:eq2}
Q_n \overset{d}{=} n(AN_n)^{-1/\gamma}, \quad \text{or} \quad N_n \overset{d}{=} (Q_n/n)^{-\gamma}/A,
\end{equation}
with the equality indicating equality in distribution.

In general, a closed form analytic expression for $G_\gamma(x)$ is not known, with the exception of the case $\gamma=\frac{1}{2}$ (corresponding to $z=3$)
\cite{levy_sur_1940, feller_introduction_1968, klages_anomalous_2008}
\begin{equation}
G_{1/2}(x) = 2 \left\{ 1-\Phi\left[ \left( \frac{\pi}{2x} \right)^{\frac12} \right] \right\}
\end{equation}
for $\Phi(x)$ the CDF of a standard normal random variable,
\begin{equation}
\Phi(x) = \frac12 \left[ 1 + \erf(x/\sqrt2) \right].
\end{equation}
The corresponding probability density function (PDF) is
\begin{equation} \label{eq:eq3}
p_{Q_n}(x) = \frac12 x^{-3/2} \exp(-\frac{\pi}{4x}).
\end{equation}
Thus we can use a simple transformation of random variables to obtain a PDF for $N_n$: if $N_n = g(Q_n)$, then
\begin{align}
p_{N_n}(x) &= p_{Q_n}(g^{-1}(x)) \abs{\dv{x}g^{-1}(x)} \nonumber \\
&= \frac12 (n(Ax)^{-2})^{-3/2} \exp(-\frac{\pi}{4} \left[ n(Ax)^{-2} \right]^{-1}) \abs{-2nA^{-2}x^{-3}} \nonumber \\
&= A n^{-1/2} \exp(-\frac{A^2\pi}{4n} x^2 ), \qquad x>0
\end{align}
which we find to be of half-Gaussian form. Integrating over $x>0$ we find the result is indeed normalised to $1$, and we can also show (via a trivial inspection) that the expectation,
\begin{align}
\Exp[N_n] &= \int_0^\infty Axn^{-1/2} \exp(-\frac{A^2\pi}{4n}x^2) \dd{x} \nonumber \\
&= \left[ \frac{-2}{A\pi} n^{1/2} \exp(-\frac{A^2\pi}{4n}x^2) \right]_0^\infty = \frac{2}{A\pi}n^{1/2}
\end{align}
grows at the correct rate $n^\gamma$, as we expect for the closed system; this result was indicated using this method in the general case in \cite{gaspard_sporadicity_1988} and numerically verified in \cite{korabel_deterministic_2004}.

\subsection{Open Gaspard-Wang model\label{sec:op-gw}}

The key idea for the open system is as follows: in \cite{gaspard_sporadicity_1988} it is argued that the Lyapunov stretching in the closed system is directly proportional, asymptotically, to $N_n$. The asymptotic qualities of $N_n$ in the closed case are shown above, and in \cite{gaspard_sporadicity_1988}. In the open case we maintain the same reasoning applies, but the distribution of times between reinjections changes in a fairly subtle way: namely, since each reinjection carries a definite risk of escape (with some probability, which we call $\ve$), our sample, which by necessity is of those trajectories which \emph{do not} escape, is biased in favour of trajectories with fewer reinjections -- geometrically with the number of reinjections -- and therefore longer excursions to the fixed point. We represent this directly with a very simple transformation: if $N_n$ is the number of reinjections in $n$ timesteps of the \emph{closed} process, let $N_n^*$ be the number of reinjections in $n$ timesteps of the \emph{open} process, with respective density functions $p_{N_n}(x)$ and $p_{N^*_n}(x)$; then
\begin{equation} \label{eq:eq4}
p_{N^*_n}(x) \propto p_{N_n}(x) \cdot (1-\ve)^x = p_{N_n}(x) \cdot \exp(-\nu x)
\end{equation}
for $\nu := -\ln(1-\ve)$, with a suitable renormalisation. Put differently, if every reinjection carries a risk $\ve$ of escape, then the distribution of $N_n^*$ will look like $N_n$, except that each trajectory is `de-weighted' by $\ve$ with each reinjection it experiences.

For simplicity we work directly with the case $\gamma = \frac12$. For the open system, we have
\begin{equation} \label{eq:opening}
p_{N_n^*}(x) \propto p_{N_n}(x) \cdot (1-\ve)^x = A n^{-1/2} \exp(-\frac{A^2 \pi}{4n} x^2 - \nu x )
\end{equation}
whose right-hand-side integrates to
\begin{equation}
B(n) = \exp(\frac{\nu^2 n}{A^2 \pi})\erfc\left(\sqrt{\frac{\nu^2 n}{A^2 \pi}}\right).
\end{equation}
This function has the Puiseux expansion at $n=\infty$
\begin{equation}
\exp(\xi)\erfc(\sqrt{\xi}) = \frac{1}{\sqrt{\pi}} \left( \xi^{-1/2} - \frac12 \xi^{-3/2} + \frac34 \xi^{-5/2} - \frac{15}{8} \xi^{-7/2} + \dots \right)
\end{equation}
with $\xi = \nu^2 n / A^2 \pi$. We remark that in practice, direct computer calculations of $B(n)$ tend to fail for large $n$ due to numerical precision problems, namely that that $\exp(\xi)$ grows very large with $\xi$ while
$\erfc(\sqrt{\xi})$ becomes very small in almost exact inverse proportion. Therefore we substitute the above Puiseux series, which we found to match well with relatively few terms to machine calculations. We also remark that by construction, the normalising function $B(n)$ gives the proportion of trajectories remaining in the ensemble at time $n$, which is of dynamical interest to us independently of $N_n^*$; it effectively represents a survival probability which be used to calculate the distribution of escape times, and thus to define escape rates.

The expectation of $N_n^*$ can be calculated to be
\begin{align}
\Exp[N_n^*] &= \int_0^\infty \frac{Axn^{-1/2}}{B(n)} \exp(-\frac{A^2\pi}{4n} x^2 - \nu x) \dd{x} \nonumber \\
&= \frac{2n^{1/2}}{A \pi B(n)} - \frac{2n\nu}{A^2\pi}.
\end{align}
In order to establish the asymptotic behaviour of the calculated expectation, we expand the first term as a Puiseux series:
\begin{align}
[B(n)]^{-1} &= \pi^{1/2} \left[ \frac{A \pi^{1/2}}{\nu n^{1/2}} - \frac{A^3\pi^{3/2}}{2\nu^3n^{3/2}} + \dots \right]^{-1} \nonumber \\
&=  \frac{\nu n^{1/2}}{A} \left[ 1-\frac{A^2\pi}{2\nu^2n}+\dots \right]^{-1} \nonumber \\
&= \frac{\nu n^{1/2}}{A} + \frac{A\pi}{2\nu n^{1/2}} + \dots
\end{align}
and therefore
\begin{align}
\Exp[N_n^*] &= \frac{2n^{1/2}}{A \pi}\left[ \frac{\nu n^{1/2}}{A} + \frac{A\pi}{2\nu n^{1/2}} + \dots \right] - \frac{2n\nu}{A^2\pi} \nonumber \\
&= \frac{2n\nu}{A^2\pi}-\frac{2n\nu}{A^2\pi} + \frac{1}{\nu} + \dots \simeq \frac{1}{\nu}
\end{align}
ie.\ in the open case, the expectation of $N_n^*$ is bounded as $n\to\infty$ and converges to $1/\nu$. This can also be seen by directly evaluating the limiting density of $N_n^*$ as $n\to\infty$,
\begin{equation} \label{eq:eq5}
p_{N_n^*} = \frac{An^{-1/2}}{B(n)} \exp(-\frac{A^2\pi}{4n} x^2 - \nu x) \to \nu \exp(-\nu x)
\end{equation}
as $n\to\infty$, by taking $B(n) = \nu n^{1/2} / A$ as above. Thus we see the distribution of $N_n^*$ converges to an exponential distribution with mean $1/\nu$, which is of course consistent with our more detailed calculations above.

\section{Average stretching per reinjection}

Now that we have our model to determine $\langle N_n^* \rangle$, it remains (for the purposes of comparison with numerical results) to derive from this an approximation to $\langle \Lambda_n \rangle$. We take inspiration from \cite{gaspard_sporadicity_1988} which argues (wrt.\ the closed system) these should be proportional by a constant representing the average stretching incurred per reinjection event.

We begin by noting that there is a one-to-one decreasing correspondence between the location of a particle in the interval $x$ (near the fixed point), and the time until its escape, $\te(x)$, derived from the differential equation for the laminar motion near the fixed point, given by \cite{korabel_fractal_2007, pegler_anomalous_2017, brevitt_singularity_2024}
\begin{equation}
\dv{x}{t} = ax^z \implies \te(x) = \gamma(x^{-1/\gamma}-1)/a, \quad \text{or} \quad x = [a\te/\gamma+1]^{-\gamma}.
\end{equation}
The instantaneous rate of Lyapunov stretching experienced by a particle at a point $x$ in its laminar phase is given by
\begin{align}
\ln\abs{M'(x)} &= \ln(1+azx^{z-1}) = \ln(1+a\frac{\gamma+1}{\gamma}x^{1/\gamma}) \nonumber \\
&= \ln(1+\frac{a(\gamma+1)}{\gamma+a\te}) \approx \frac{a(\gamma+1)}{\gamma+a\te}. \label{eq:lnt}
\end{align}
Thus the total Lyapunov stretching endured by a particle over the duration of an excursion of length $\tau$ is
\begin{equation}
I(\tau) = \int_0^\tau \ln(1+\frac{a(\gamma+1)}{\gamma+a\te}) \dd{\te}
\end{equation}
which has an exact closed form expression which is asymptotically
\begin{equation}
I(\tau) \simeq (\gamma+1)\ln(\tau) + C
\end{equation}
whose constant $C$ can be numerically estimated.

Gaspard and Wang \cite{gaspard_sporadicity_1988} reasoned that since each escape/reinjection event necessitates a certain amount of stretching in accordance with its duration, we may obtain $\Exp[\Lambda_n] \simeq C\Exp[N_n]$ for some bounded constant $C>0$. Following along these lines we obtain the naïve constant factor
\begin{equation} \label{eq:ibar}
\bar{I} = \Exp_w I = \int_0^\infty I(\tau) w(\tau) \dd{\tau}.
\end{equation}
This integral is expected to converge, albeit slowly; while $w(\tau)$ has power law tails, $I(\tau)$ is asymptotically logarithmic in $\tau$, meaning the tail of the integrand decays as $\tau^{-\gamma-1}\ln\tau$. Since this slow convergence is impractical for numerical integration, we may instead transform \eqref{eq:ibar} by a complete change of variable \cite{ross_introduction_2014}, to
\begin{equation}
\Exp_w I = \int_0^\infty w(I^{-1}(\tau)) (I^{-1})'(\tau) \dd{\tau}.
\end{equation}
This is equivalent to the direct expectation $\Exp [I(T)] = \int I(t)\,p_T(t) \dd{t}$ for a random variable $T$ with density function $w(\tau)$. The tail of this integrand is approximately exponential, which allows for easier numerical integration. It was by this method that we obtained $\bar{I}$ to fit our model $\langle \Lambda_n \rangle \simeq \bar{I} \langle N_n \rangle$.

However we may also opt for a more nuanced formula; for example, we may reason that $T$ cannot exceed the total time ($T\leq n$), and thus we can restrict the upper bound of the integral
\begin{equation} \label{eq:rainers}
\bar{I}(n) = \Exp_w [ I(\tau) \,|\, \tau < n ] = \int_0^n I(\tau) \tilde{w}_n (\tau) \dd{t}, \quad \tilde{w}_n (\tau) := \frac{w(t)}{\int_0^n w(t) \dd{t}};
\end{equation}
this is the variant we ultimately chose to use most extensively when testing against simulations. However we remark that, as $n\to\infty$, $\bar{I}(n) \to \bar{I}$ as above.

Testing these models against the observed numerical results from simulations, we see a generally good qualitative fit. Indeed, the smoother transient of \eqref{eq:rainers} seems to replicate the simulations very well, much better for smaller $n$ than \eqref{eq:ibar}.

In order to calculate the stretching from an incomplete segment of a trajectory (ie.\ a segment consisting of a motion from $A_k$ to $A_m$, for $k>m$, without visiting $A_0$), we simply observe that the formulas $I(t)$ discussed above measure the stretching \emph{backwards} in $t$ from the time of reaching $A_0$, and thus if we have an incomplete segment beginning at time $t_1$, and ending at time $t_2$, we may calculate the time at which this trajectory {would} eventually reach $A_0$ by
\begin{equation}
t_c = t_1 + \te(x_{t_1}) = t_2 + \te(x_{t_2})
\end{equation}
and thus deduce the stretching over the incomplete phase $[t_1, t_2]$ to be
\begin{align}
I_\mathrm{incomplete}[t_1, t_2] &= I_\mathrm{complete}[t_1,t_c] - I_\mathrm{complete}[t_2,t_c] \nonumber \\
&= I(t_c - t_1) - I(t_c - t_2)
\end{align}
using the formulae above. Readers are referred to the diagram Fig.~4 in the main paper, which we do not reproduce here.

\section{Contribution from incomplete segments}

In this section we show the calculation of the contribution from the incomplete segments of trajectories, which manifest as `ticks' towards the end of the measurement as $n\to t$. To begin to get an understanding of the observed `ticks', we must first prove a few general statements about probability.

Let $X$ be a random variable obeying a generic Pareto Type II random variable with shape parameter $\alpha$ and generic location and scale parameters $\mu,\sigma$, notated $X \sim \Pareto(\alpha, \mu, \sigma)$ with density function
\begin{equation}
f_X(t) = f^\mathrm{Pareto}_X(t; \alpha, \mu, \sigma) = \begin{cases} \frac{\alpha}{\sigma} \left( 1 + \frac{t-\mu}{\sigma} \right)^{-1-\alpha}, &t\geq\mu;\\ 0,&\mathrm{otherwise} \end{cases}
\end{equation}
Then if $Y$ is such that $\Prb[Y\in\cdot\,] = \Prb[X\in\cdot \;|\, X \geq \mu + A]$,
\begin{equation}
f_Y(t) = \begin{cases} \frac{f_X(t)}{\int_{\mu+A}^\infty f_X(\tau) \dd{\tau}}, &t\geq\mu+A;\\ 0,&\mathrm{otherwise} \end{cases}
\end{equation}
with the density in the top line given by
\begin{align}
\frac{f_X(t)}{\int_{\mu+A}^\infty f_X(\tau) \dd{\tau}}
 &= \frac{\frac{\alpha}{\sigma} \left( 1 + \frac{t-\mu}{\sigma} \right)^{-1-\alpha}}{\left( 1+\frac{A}{\sigma} \right)^{-\alpha}}
 = \frac{\alpha}{\sigma \left( 1+\frac{A}{\sigma} \right)} \left( \frac{1+\frac{t-\mu}{\sigma}}{1+\frac{A}{\sigma}} \right)^{-1-\alpha} \nonumber \\
 &= \frac{\alpha}{\sigma+A} \left( \frac{\sigma+t-\mu}{\sigma+A} \right)^{-1-\alpha}
 = \frac{\alpha}{\sigma+A} \left( 1 + \frac{t-\mu-A}{\sigma+A} \right)^{-1-\alpha} \nonumber \\
 &= f^\mathrm{Pareto}_X(t; \alpha, \mu+A, \sigma+A).
\end{align}
Therefore if $X \sim \Pareto(\alpha, \mu, \sigma)$ with density $f_X(t; \alpha, \mu, \sigma)$, then $X$ conditioned on $X \geq \mu + A$ has density $f^\mathrm{Pareto}_X(t; \alpha, \mu+A, \sigma+A)$.

Note also that $\mu$ and $\sigma$ form a location-scale parameter space for $\Pareto(\alpha, \mu, \sigma)$; ie.\ if $X \sim \Pareto(\alpha, 0, 1)$, then $\sigma' X + \mu' \sim \Pareto(\alpha, \mu', \sigma')$.

In our case,
\begin{equation}
w(t) = \frac{\gamma b^\gamma}{(b+t)^{1+\gamma}}
 = \frac{\frac{\gamma}{b}}{(1+\frac{t}{b})^{1+\gamma}}
 = f^\mathrm{Pareto}_X(t; \gamma, 0, b)
\end{equation}
and so if $f_X(t) = w(t)$, then we may write that $Y$ satisfying $f_Y(t) = f_X(t \,|\, X \geq A)$ has $Y \overset{d}{=} \frac{b+A}{b} X + A$ {in distribution}. For large $A$, this means both
\begin{equation}
Y \overset{d}{\sim} \frac{A}{b} X + A \quad \mathrm{and} \quad Y-A \overset{d}{\sim} \frac{A}{b} X
\end{equation}
\emph{asymptotically} {in distribution}, which both scale linearly with $A$.


The quantity $m$, labelled as the `overhang' in the diagram Fig.~4 in the main paper, measures the time between the end of the measurement time $t$ and the time at which the incomplete segment -- the segment in progress as the measurement time in reached -- \emph{would have} been reinjected (exited the left-hand branch). This incomplete segment begins at the time of the final reinjection event, which we label $\tau$. Then we can deduce a probability distribution for $m$ by
\begin{align} \label{eq:eq35}
f_m(t') = w(t' \,|\, t' > t-\tau) &= f^\mathrm{Pareto}_X \left( t'; \gamma, 0, \frac{b+t-\tau}{b} \right) \nonumber \\
&= \frac{\gamma b}{b+t-\tau} \left( 1 + \frac{t'b}{b+t-\tau} \right)^{-1-\gamma}
\end{align}
and the time until escape for a particle in the incomplete phase is given by
\begin{equation} \label{eq:eq36}
t_\mathrm{esc}(t') = t+m-t'.
\end{equation}
Finally, the instantaneous rate of stretching $\ln\abs{M'(x_{t'})}$ at a time $t'$ in the incomplete phase can be evaluated from \eqref{eq:lnt} to be
\begin{equation} \label{eq:eq37}
\lambda(t') = \frac{a(\gamma+1)}{\gamma+t_\mathrm{esc}} = \frac{a(\gamma+1)}{\gamma+t+m-t'}.
\end{equation}

Then, defining $\Lambda_t^\mathrm{inc}(t')$ as the average cumulative stretching in the incomplete phase up to time $t'<t$, given a measurement time $t$, we calculate by averaging over $m$
\begin{equation}
\dv{t'}\Lambda_t^\mathrm{inc}(t') = \Exp_{m}[\lambda(t')].
\end{equation}
We evaluate this expectation below: taking $y = \frac{\lambda(t')}{a(\gamma+1)}$, and $g$ to be the transformation $m \mapsto y$,
\begin{equation}
y=g(m)=\frac{1}{\gamma+t+m-t'},
\end{equation}
we evaluate
\begin{equation}
\Exp_{m}[\lambda(t')] = a(\gamma+1) \int_{\R} y\,f_y(y) \dd{y}.
\end{equation}
To determine the density $f_y(y)$, we perform a standard change of variables: we have $m=g^{-1}(y) = 1/y -t+t'-\gamma$ and so we take
\begin{align}
f_y(y) &= f_m(g^{-1}(y)) \abs{\dv{y} g^{-1}(y)} \nonumber \\
 &= \frac{\gamma b}{b+t-\tau} \left( 1 + \frac{(1/y-\gamma-t+t')b}{b+t-\tau} \right)^{-1-\gamma} \frac{1}{y^2}.
\end{align}
If $m \in [0,\infty)$ then $y \in (0, \frac{1}{\gamma+t-t'}]$ and so
\begin{equation}
\Exp_{m}[\lambda(t')] = a(\gamma+1) \int_0^{\frac{1}{\gamma+t-t'}}
\frac{\gamma b}{(b+t-\tau)y} \left( 1 + \frac{(1/y-\gamma-t+t')b}{b+t-\tau} \right)^{-1-\gamma}.
\end{equation}

We want to evaluate this expectation for all $t' \in [\tau,t]$ (between the final reinjection event and the end of the measurement time), crucially in the shape and scaling of the solution in the limit $t\to\infty$. We also remark that the time of the final reinjection $\tau$ is also a random variable we will need to consider when fitting these ticks into their proper context.

From this analysis we may immediately remark by inspection that the time quantities $m$ and $t_\mathrm{esc}$ scale in proportion with $t-\tau$, see \eqref{eq:eq35} and \eqref{eq:eq36}, while the stretching $\lambda(t')$ scales in proportion with $\frac{1}{t-\tau}$, see \eqref{eq:eq37}. Further, since the total stretching across the incomplete period is given by
\begin{equation}
\Lambda_t^\mathrm{inc}(t) = \int_\tau^t \lambda(t') \dd{t'}
\end{equation}
we see that, by combination of the above, $\Lambda_t^\mathrm{inc}(t)$ is of order $\Orb(1)$ with respect to $t-t'$. This is of course also what we observed numerically. In order to deal with the scaling of these various quantities, and produce the shape of the `ticks', we can `non-dimensionalise' according the following scheme:
\begin{itemize}
\item We change our frame of reference to fix $0$ at $\tau$ and $1$ at $t$. This shrinks the space by a factor of $t-\tau$.
\item We fix $\varepsilon := \frac{1}{t-\tau}$. Thus the limit $t\gg \tau$ corresponds to the limit $\ve\to 0$ -- we will look at the dependence of $\tau$ on $t$ later. We then have:
\begin{itemize}[label={\tiny$\blacksquare$}]
\item $\tilde{m} = \frac{m}{t-\tau} \sim \Pareto(t'; \gamma, 0, \frac{1}{b}+\ve)$;
\item $\tilde{t'} = \frac{t'-\tau}{t-\tau} \in [0,1]$;
\item $\tilde{t}_\mathrm{esc} = \frac{t_\mathrm{esc}}{t-\tau} = 1-\tilde{t'}+\tilde{m}$;
\item $\tilde{\lambda}(\tilde{t'}) = (t-\tau)\lambda(t') = \frac{a(\gamma+1)}{\gamma\ve+1-\tilde{t'}+\tilde{m}}$;
\item $\tilde{\Lambda}^\mathrm{inc}(\tilde{t'}) = \Lambda_t^\mathrm{inc}(t') = \int_0^{\tilde{t'}} \tilde{\lambda}(\tilde{t'}) \dd{\tilde{t'}}$.
\end{itemize}
\end{itemize}
Our integral can then become
\begin{align}
\dv{\tilde{t'}} \tilde{\Lambda}^\mathrm{inc}(\tilde{t'}) &= \Exp_{\tilde{m}}[\tilde{\lambda}(\tilde{t'})] \nonumber \\
&= a(\gamma+1) \int_0^{\frac{1}{\gamma\ve+1-\tilde{t'}}} y \, \frac{\gamma}{1/b + \ve} \left( 1 + \frac{1/y-\gamma\ve-1+\tilde{t'}}{1/b + \ve} \right)^{-1-\gamma} \frac{1}{y^2} \dd{y}.
\end{align}
In the limit $\ve\to 0$ this becomes
\begin{equation}
\Exp_{\tilde{m}}[\tilde{\lambda}(\tilde{t'})] = a(\gamma+1) \int_0^{\frac{1}{1-\tilde{t'}}} \frac{\gamma b}{y} \left( 1 + (1/y-1+\tilde{t'})b \right)^{-1-\gamma} \dd{y}
\end{equation}
which evaluates (with the aid of Wolfram Mathematica) to
\begin{equation}
\dv{\tilde{t'}} \tilde{\Lambda}^\mathrm{inc}(\tilde{t'})
 = \frac{a \gamma}{b^\gamma (1-\tilde{t'})^{\gamma+1}} \; {}_2F_1 \left( 1+\gamma,1+\gamma;2+\gamma; 1 - \frac{1}{b(1-\tilde{t'})} \right)
\end{equation}
where ${}_2F_1(a,b;c;z)$ is Gauss' hypergeometric function. This can be evaluated for all $\tilde{t'}\in [0,1)$, and has a singularity at $\tilde{t'}=1$ when $\ve=0$. This singularity can be shown to be logarithmic, and therefore integrable: according to the integral representation given in \cite{bateman_higher_1953}, the asymptotics of the hypergeometric function ${}_2F_1(1+\gamma,1+\gamma; 2+\gamma; z)$ for $z \notin [1,\infty)$ are given by
\begin{equation}
(\gamma+1)(-z)^{-1-\gamma} \left[ \log(-z)+C+\Orb(z^{-1}) \right]
\end{equation}
and therefore allowing $x = 1-\tilde{t'}$,
\begin{align}
\dv{t'}\tilde{\Lambda}^\mathrm{inc}(\tilde{t'}) &\simeq \frac{a\gamma}{b^\gamma x^{1+\gamma}} \left( (\gamma+1) \left( \frac{1}{bx}-1 \right)^{-1-\gamma} \log(\frac{1}{bx}-1) \right) \nonumber \\
&\simeq a \gamma (\gamma+1) b^{-\gamma} x^{-1-\gamma} \left( \frac{1}{bx} \right)^{-1-\gamma} \log(\frac{1}{bx}) \nonumber \\
&\simeq a \gamma (\gamma+1) b \log(\frac{1}{bx})
\end{align}
for $x \to 0_+$. We also find that both of the above formulae correspond well to results from numerical integration for small but non-zero $\ve$. Therefore the desired quantity
\begin{equation}
\tilde{\Lambda}^\mathrm{inc}(\tilde{t}') = \int_0^{\tilde{t}'} \Exp_{\tilde{m}}[\tilde{\lambda}(t^\dagger)] \dd{t^\dagger}
\end{equation}
is well-defined and finite.

We remark that $\tilde{\Lambda}^\mathrm{inc}(\tilde{t}')$ is a function defined on $0 \leq \tilde{t'} \leq 1$ independent of $t$, and that, returning to the definitions, the desired quantity $\Lambda_t^\mathrm{inc}(t')$ is given by
\begin{equation}
\Lambda_t^\mathrm{inc}(t') = \tilde{\Lambda}^\mathrm{inc}(\tilde{t}') = \tilde{\Lambda}^\mathrm{inc} \left( \frac{t'-\tau}{t-\tau} \right)
\end{equation}
for $\tau \leq t' \leq t$. In principle, $\tau$ is random in the range $0<\tau<t$, and its distribution is dependent on $t$, and to obtain a suitable function $\Lambda_t^\mathrm{inc}(t')$ we should average over $\tau$; however we remark that, especially for large $t$, typically $\tau \ll t$, and so for simplicity we heuristically take
\begin{equation}
\Lambda_t^\mathrm{inc}(t') = \tilde{\Lambda}^\mathrm{inc}(\tilde{t}') \approx \tilde{\Lambda}^\mathrm{inc}(t'/t)
\end{equation} 
for $0\leq t'\leq t$. Adding $\Lambda_t^\mathrm{inc}(t')$ onto $\Exp[N_n^*]$ for varying values of $t$, as shown in the main paper, produces ticks that look qualitatively very similar to those observed in simulations.

\section{Reconstruction of the fractal repeller}

By visual inspection from the numerical measure, we make the observation that if $x_n$ is a point in the neighbourhood of a preimage (of some order $k$) of $x=0$, and $M(x_{n-1}) = x_n$, then $x_{n-1}$ is also a point in the neighbourhood of a preimage (of order $k+1$) of the fixed point, and
\begin{equation} \label{eq:cumhistlaw}
\mu(\dd x_{n-1}) \approx \frac{1}{\abs{M'(x_{n-1})}}\,\mu(\dd x_n).
\end{equation}
To demonstrate this, we construct a $k^\mathrm{th}$-order approximation of the repeller as follows: a $k^\mathrm{th}$-order preimage $p^x_{i,k}$ of a point $x$ is a point such that $M^k(p^x_{i,k})=x$.
We label as $\{p^0_{i,k}\}$ the set of all $k^\mathrm{th}$-order preimages of the point $x=0$, indexed by $i$. We assign to each point $p^0_{i,k}$ a weight $0<w_{i,k}<1$ according to \eqref{eq:cumhistlaw}, ie., if $M(p^0_{i,k+1})=p^0_{j,k}$, then
\begin{equation}
w_{i,k+1} = \frac{1}{\abs{M'(p^0_{i,k+1})}}\,w_{j,k},
\end{equation}
assigning to $x=0$ the initial weight $w_{i,0}=1$, so that we obtain
\begin{equation}
w_{i,k} = \left( \prod_{k'=0}^{k-1} \abs{(M^{(k')})'(p^0_{i,k})} \right)^{-1}.
\end{equation}
The $k^\mathrm{th}$-order approximation of the repeller $\mu_k$ is then given by the normalised weighted sum of delta functions on these preimages:
\begin{equation}
\mu_k = \frac{\sum_{\{p^0_{i,k'}\,:\,k'\leq k\}} w_{i,k'}\,\delta(p^0_{i,k'}) }{\sum_{\{p^0_{i,k'}\,:\,k'\leq k\}} w_{i,k'} }.
\end{equation}
In the corresponding figure in the main paper, we show the CDFs of these approximations $\mu_k$ for increasing values of $k$. It is evident that these approximations converge from above onto the measure generated by the numerics, which converge from below. We hypothesise that the `true' fractal repeller is exactly the limiting measure of both of these measure sequences.



\section*{References}
\bibliographystyle{iopart-num}
{
\bibliography{bib_total}
}

%
%